\documentclass[twocolumn,showpacs,amsmath,amssymb,prb]{revtex4}

\usepackage{epsfig}
           
\usepackage{dcolumn}
     
\usepackage{bm}

\usepackage{amsmath}

\usepackage{amsfonts}

\usepackage{amssymb}

\usepackage{graphicx}%

\newcommand{\beq}{\begin{eqnarray}}

\newcommand{\eeq}{\end{eqnarray}}

\renewcommand{\vec}[1]{{\mathbf{#1}}}

\begin{document}

\title{Dictionary between scattering matrix and Keldysh formalisms
for quantum transport driven by time-periodic fields.}  

\author{Liliana Arrachea$^{(1), (2)}$ and Michael Moskalets$^{(3)}$.}
\affiliation{$^{(1)}$ Departamento de F\'{\i}sica de la Materia Condensada,
Universidad de Zaragoza, Pedro Cerbuna 12 (50009) Zaragoza, Spain.\\
$^{(2)}$ Instituto de Biocomputaci\'on y F\'{\i}sica de Sistemas
  Complejos, Universidad de Zaragoza,
Corona de Arag\'on 42, (50009) Zaragoza, Spain.\\
$^{(3)}$ Department of Metal and Semiconductor Physics, National Technical
University, ``Kharkov Polytechnical Institute'', (61002) Kharkov, Ukraine.
}


\begin{abstract}
We present  the relation between the Floquet scattering matrix and the
 non-equilibrium Green's function formalisms to
 transport theory in noninteracting electronic systems in contact to
 reservoirs and driven by time-periodic 
fields. 
We present a translation formula that expresses the Floquet scattering matrix
 in terms of a Fourier transform of the retarded 
Green's function. 
We prove that such representation satisfies the  fundamental  identities of transport theory. 
We also present the ``adiabatic'' approximation to the dc-current in the
 language of the Keldysh formalism. 
\end{abstract}

\maketitle

\section{Introduction}
In the last years,
there has been an increasing theoretical and experimental
 activity around quantum transport phenomena
induced by time-dependent fields. 
Pumping phenomena in mesoscopic systems constitutes 
a very interesting case,
where periodic out-of-phase potentials deform the gates of semiconductor
structures allowing for the generation of dc-currents even in the absence
of a static bias. 
\cite{SMCG99,GAHMPEUD90,DMH03,VDM05,BTP94,Brouwer98,AA98,ZSA99,AEGS00,
EWAL02,PB01,MCM02,AEGS04,ZLCMcK04}

The scattering matrix approach and Keldysh non-equilibrium Green's function
technique are the most powerful formalisms in the theory of quantum transport.
Recently, the generalization of the Scattering (S) Matrix 
theory to time-periodic transport phenomena in an energy representation has
been formulated \cite{micha1,micha2,micha3}, while
an alternative treatment in a time representation has been proposed in
Refs. ~\onlinecite{VAA01,PB03}.
Keldysh formalism as a theoretical tool to investigate time-dependent
transport phenomena in mesoscopic systems 
has
been introduced some time ago \cite{past,JWM94} and has been employed to
study problems like ac-transport through quantum dots \cite{plat1} and superlattices
\cite{plat2}, Josephson junctions \cite{joseph1,joseph2} and quantum pumps \cite{pumpkel1,pumpkel2}.
Recently a practical formulation  to treat problems with harmonically
time-dependent potentials has been presented and used  to investigate quantum transport
in a mesoscopic ring threaded by a time-dependent flux \cite{lilir} and systems
with ac-potentials \cite{lilip,lilire}. 
Other formalisms to describe quantum transport in the presence of time-periodic
fields are based in a modified transfer matrix approach \cite{sols} and in
 the Floquet representation of the Hamiltonian with the
introduction of non-hermitian dynamics for the  wave function propagation,
in order to represent dissipative effects. \cite{flohan,foa,kim}

The Scattering Matrix formalism is basically a single-particle
approach. Therefore,
it cannot be directly applied to systems described by Hamiltonians
containing
many-particle interactions. The theoretical framework in
which Keldysh formalism is based is exactly the opposite one,
namely the systematic treatment of many-particle interacting systems.
This formalism is, however, also adequate to investigate transport phenomena through
mesoscopic systems even in the case that many-body interactions do not play
a relevant role. The reason it that  the effect of
the environment, in particular, the leads and reservoirs to which the
mesoscopic system is connected, are suitably represented in terms of self-energies.
In the description of  quantum transport in systems of non-interacting electrons,
the agreement between both
formalisms is expected to be
the rule.  
In the context of stationary transport, the equivalence between the two approaches 
was first pointed out by Fisher and Lee. \cite{file}

An important experimental situation corresponds to the case of
slowly oscillating driving fields.
The low frequency regime is sometimes loosely referred to as 
``adiabatic''. This word stems from the Greek word ``a-diabatos'', which 
means ``not passable''. Traditionally, in 
theoretical physics, this term is employed
when an isolated or closed quantum mechanical system
is perturbed 
by a time-dependent Hamiltonian in such a way that the eigenstates 
do not mix as time evolves. This idea cannot be trivially
exported to describe  a quantum
system coupled to an environment where the spectrum is continuum and
concepts like energy levels are not well defined.
In the framework of open quantum systems,
 the term ``adiabatic'' 
 is sometimes understood as a synonymous of low frequency behavior
while it is also sometimes employed to
define a description where the
variable $t$ in the time-dependent piece of the Hamiltonian is considered as a 
frozen parameter. An important number of works have been devoted to
investigate quantum transport in pumps within the low-frequency regime.
In particular, an ``adiabatic'' approximation to the Floquet scattering matrix has
been introduced \cite{BTP94,Brouwer98,AEGS04,micha1, micha2, micha3} which, in practice,
allows for the evaluation of the contribution to the pumped dc-current that behaves linearly in the driving frequency. 

The aim of this work is twofold. On one hand, we
 show that, 
for non-interacting quantum systems
driven by time-periodic fields, in contact to
static reservoirs with arbitrary densities of states, or with
oscillating reservoirs described by smooth densities of states,
it is possible to establish a transparent and complete dictionary
between the Floquet scattering matrix approach of Refs.~\onlinecite{micha1,micha2,micha3}
 and the Keldysh Green's function treatment of Ref.~\onlinecite{lilip}. 
The second goal of this work is to 
formulate an ``adiabatic'' approximation, analogous to the one used in
scattering matrix formalism,
in the language of  nonequilibrium Green's functions.

The work is organized as follows. In section II we summarize the description
of quantum transport within the scattering matrix approach as well as the
``adiabatic'' approximation to the Floquet scattering matrix. The description
of
quantum transport for non-interacting electrons driven by time-periodic fields
in the framework of Keldysh formalism
is summarized in section III. We present several equivalent equations to
evaluate the dc-component of the current flowing between the reservoirs and
the central mesoscopic driven system. We consider cases where the pumping voltages are
locally applied only at the central system, as well as cases where the voltages are
applied at the reservoirs.
 A formula that allows for the translation
between the two formalisms is presented in section IV. In section V we show
that a fundamental property like 
the unitarity of the Floquet S-matrix can be proved in
terms of identities satisfied by the non-equilibrium Green's functions. In
section VI we present the adiabatic approximation formulated in the
language  of non-equilibrium Green's functions. Finally, section VII is devoted
to
summary and conclusions. We have also included appendices A, B, C, D and E 
with relevant identities and properties used along the work.

\section{Scattering matrix formalism}
\label{smf}

\subsection{General formalism}
The scattering matrix approach to quantum transport, usually referred to as 
the Landauer-B\"uttiker approach, \cite{Landauer75,Buttiker92} considers the
propagation of carriers through a mesoscopic sample as a 
scattering process. 
The sample (scatterer) is assumed to be connected to several $N_r$ contacts
(playing the role of electron reservoirs) via single 
channel leads. 
Then, all the information about transport properties of a mesoscopic sample is
encoded in the scattering matrix $\hat S$ 
whose elements are quantum mechanical amplitudes for electrons
coming from some lead to be scattered into the same or any other lead.
These amplitudes are normalized in such a way that their square define
the corresponding particle fluxes (currents).

In each lead there are two kinds of states, incoming to and outgoing off the scatterer.
Correspondingly, we introduce two kinds of second-quantization operators 
$\hat a_{\alpha}(E)$/$\hat a^{\dagger}_{\alpha}(E)$ and 
$\hat b_{\alpha}(E)$ /$\hat b^{\dagger}_{\alpha}(E)$
which annihilate/create one particle in the lead $\alpha$ with energy $E$. 
Using these operators one can calculate a time-dependent current flowing into
lead $\alpha$ as follows (we use a convention 
that the current directed from the scatterer towards the reservoir is positive): \cite{Buttiker92} 
\begin{equation}
\label{smf_1}
\begin{array}{c}
I_{\alpha}(t) =  \frac{e}{h}\int\limits_{0}^{\infty}\int\limits_{0}^{\infty}dEdE^{\prime}
e^{i\frac{E-E^{\prime}}{\hbar}t} \\
\ \\
\times \left\{
\langle  \hat b^{\dagger}_{\alpha}(E)\hat b_{\alpha}(E^{\prime})\rangle 
- \langle  \hat a^{\dagger}_{\alpha}(E)\hat a_{\alpha}(E^{\prime})\rangle 
\right\}.
\end{array}
\end{equation}
Here  $\langle\dots\rangle $ means quantum-statistical averaging over
equilibrium states of all the reservoirs which are assumed 
to be unaffected by the coupling to the scatterer.

\subsubsection{Time-periodic local potentials and static reservoirs}
The particles incoming from some reservoir $\alpha$ are equilibrium particles. 
Therefore, if the reservoirs are stationary we have:
\begin{equation}
\label{smf_2}
\langle  \hat a^{\dagger}_{\alpha}(E)\hat a_{\beta}(E^{\prime})\rangle  = 
\delta_{\alpha\beta}\delta(E-E^{\prime})f_{\alpha}(E),
\end{equation}
where $f_{\alpha}(E)$ is the Fermi distribution function for electrons in reservoir $\alpha$.
In general, the reservoirs have different chemical potentials $\mu_{\alpha}$
and temperatures $T_{\alpha}$. The oscillating 
potentials at reservoirs will be considered separately.

The operators $\hat b_{\alpha}$ for particles outgoing (i.e., scattered) into
 the lead $\alpha$ are related to the operators 
 $\hat a_{\beta}$ for incoming particles $(\beta = 1,\dots, N_r)$ through the scattering matrix. \cite{Buttiker92}
In the present paper we will consider the scatterer which is driven by
 external forces that are periodic in time with period 
$\tau_0= 2\pi/\Omega_0$.
Interacting with such a scatterer an electron can gain or loss some energy quanta $n\hbar\Omega_0$, $n=0,\pm 1,\dots$
Therefore, in this case the elements $S_{F,\alpha\beta}(E_n,E)$ of the
 scattering matrix (the Floquet scattering matrix 
$\hat S_F$, see, e.g., Ref.\onlinecite{PA04}) are photon-assisted amplitudes
 (times $\sqrt{k_n/k}$, $k=\sqrt{2mE/\hbar^2}$)
 for an electron with energy $E$ entering the scatterer through lead $\beta$
 and to leave the scatterer with energy
 $E_n = E + n\hbar\Omega_0$ through lead $\alpha$. 
Then the desired relation between operators $\hat b$ for outgoing particles
 and $\hat a$ for incoming particles reads as follows: 
\cite{micha1}
\begin{equation}
\label{smf_3}
\hat b_{\alpha}(E) = \sum\limits^{N_r}_{\beta=1}\sum\limits_{n} S_{F,\alpha\beta}(E,E_n)
\hat a_{\beta}(E_n).
\end{equation}
The sum over $n$ runs over those $n$ for which $E_n > 0$. 
In general, some elements of the Floquet scattering matrix describe
transitions between the  bound states ($E_n < 0$) 
and propagating (i.e., current-carrying) states or vice versa. 
Such processes do not contribute to Eq.(\ref{smf_3}).
Therefore, in what follows, by the Floquet scattering matrix we will mean the
submatrix corresponding to transitions between 
propagating states only. In fact,
if $\hbar \Omega_0 \ll E$, the sum in Eq.(\ref{smf_3}) runs over all the integers: $-\infty < n < \infty$.
Note that if the scatterer is stationary, then  only the term with $n=0$
remains non-vanishing.

Current conservation implies that the Floquet scattering matrix is a unitary matrix: 
\cite{micha1,micha2}
\begin{subequations}
\label{smf_4}
\begin{equation}
\label{smf_4A}
\sum\limits_{\beta=1}^{N_r}\sum\limits_{n=-\infty}^{\infty} 
S^{*}_{\beta\alpha}(E_n,E)S_{\beta\gamma}(E_n,E_m) = \delta_{m0}\delta_{\alpha\gamma},
\end{equation}
\begin{equation}
\label{smf_4B}
\sum\limits_{\beta=1}^{N_r}\sum\limits_{n=-\infty}^{\infty} 
S^{*}_{\alpha\beta}(E,E_n)S_{\gamma\beta}(E_m,E_n) = \delta_{m0}\delta_{\alpha\gamma}.
\end{equation}
\end{subequations}

One can easily check that the unitary conditions guarantee that the
operators for outgoing particles obey the same 
anti-commutation relations as the operators for incoming particles:
\begin{equation}
\label{smf_5}
\begin{array}{c}
 [\hat a^{\dagger}_{\alpha}(E),\hat a_{\beta}(E^{\prime})] 
 = \delta_{\alpha\beta}\delta(E-E^{\prime}), \\
\ \\
\ [ \hat b^{\dagger}_{\alpha}(E),\hat b_{\beta}(E^{\prime})] 
= \delta_{\alpha\beta}\delta(E-E^{\prime}).
\end{array}
\end{equation}

However, in contrast to incoming particles, the scattered ones are
non-equilibrium particles with the 
distribution function 
$f^{(out)}_{\alpha}(E)\delta(E-E^{\prime})\delta_{\alpha\beta} = \langle 
\hat b^{\dagger}_{\alpha}(E) \hat b^{\dagger}_{\beta}(E^{\prime}) \rangle$
being different from the Fermi distribution function: \cite{micha1}
\begin{equation}
\label{smf_6}
f^{(out)}_{\alpha}(E) = \sum\limits_{\beta=1}^{N_r}\sum\limits_{n=-\infty}^{\infty} 
\left|S_{F,\alpha\beta}(E,E_n) \right|^2 f_{\alpha}(E_n).
\end{equation}

Using Eqs.(\ref{smf_2}) and (\ref{smf_3}) one can calculate the time-dependent current 
$I_{\alpha}(t)$, Eq.(\ref{smf_1}), flowing into the lead $\alpha$. 
The dc component $I_{\alpha}$ of this current reads as follows: 
\begin{equation}
\label{smf_7}
I_{\alpha} = \frac{e}{h}\int\limits_{0}^{\infty}dE\left\{f_{\alpha}^{(out)}(E) 
- f_{\alpha}(E) \right\}.
\end{equation}
Substituting Eq.(\ref{smf_6}) into the above equation, using Eq.(\ref{smf_4B})
and making the shift $E\to E - n\hbar\Omega_0$ we 
finally get:
\begin{equation}
\label{smf_8}
\begin{array}{r}
I_{\alpha} = \frac{e}{h}\int\limits_{0}^{\infty}dE
 \sum\limits_{\beta=1}^{N_r}\sum\limits_{n=-\infty}^{\infty} 
\left|S_{F,\alpha\beta}(E_n,E) \right|^2 \\
\ \\
\times \left\{f_{\beta}(E) - f_{\alpha}(E_n) \right\}.
\end{array}
\end{equation}
This expression emphasizes that for weak driving amplitudes and small
 frequencies,
 only electrons close to the Fermi level do contribute to the current.

An alternative expression for the above current may be obtained by
 making the energy shift in $f_{\alpha}^{(out)}(E)$ only
and using Eq.(\ref{smf_4A}).  We get a dc current 
in a more usual form (as a difference of forward and back photon-assisted transmission probabilities):
\begin{equation}
\label{smf_9}
\begin{array}{r}
I_{\alpha} = \frac{e}{h}\int\limits_{0}^{\infty}dE
 \sum\limits_{\beta=1}^{N_r}\sum\limits_{n=-\infty}^{\infty}
\Big\{\left|S_{F,\alpha\beta}(E_n,E) \right|^2 f_{\beta}(E) \\ 
\ \\
- \left|S_{F,\beta\alpha}(E_n,E) \right|^2 f_{\alpha}(E) \Big\}.
\end{array}
\end{equation}

Given the above equations either (\ref{smf_8}) or (\ref{smf_9}) define a dc
 current flowing through the scatterer coupled to stationary
 equilibrium reservoirs. 

\subsubsection{Time-dependent voltages at the reservoirs}

If the reservoirs are subject to oscillating voltages with the same frequency:
\begin{equation}
\label{smf_10}
V_{\alpha}(t) = V_{\alpha}\cos(\Omega_0 t + \varphi_{\alpha}),
\end{equation}
then we proceed as follows. \cite{PB98,micha2}
Within the reservoir with uniform oscillating potential the electron wave
function has the structure of the Floquet function type:
\begin{equation}
\label{smf_11}
\begin{array}{r}
\Psi_{\alpha}(V_{\alpha},E,\vec r) = e^{-iEt/\hbar - i\hbar^{-1}
\int\limits_{-\infty}^{t} dt^{\prime} eV_{\alpha}(t')}\psi_E(\vec r) \\
\ \\
= e^{-iEt/\hbar}\psi_E(\vec r)\sum\limits_{n=-\infty}^{\infty}
J_{n}\left(\frac{eV_{\alpha}}{\hbar\Omega_0} \right)e^{-in(\Omega_0 t + \varphi_{\alpha})}.
\end{array}
\end{equation}
We introduce operators $\hat a^{\prime\dagger}(E)$/$\hat a^{\prime}(E)$ which
create/annihilate one particle in the given above 
Floquet state. 
The corresponding distribution function is the Fermi distribution function:
$\langle  \hat a^{\prime\dagger}_{\alpha}(E)\hat a^{\prime}_{\beta}(E^{\prime})\rangle  = 
\delta_{\alpha\beta}\delta(E-E^{\prime})f_{\alpha}(E)$, where  the Floquet
quasi-energy $E$ is chosen to be equal to the energy of 
the stationary state with the same spatial part $\psi_E(\vec r)$. 
This follows from the fact that the uniform oscillating potential does not
change the normalization of the wave function:
 $\int d^3r|\Psi(E)|^2  =  \int d^3r|\psi_E|^2 =1$.
As a consequence, the only occupied states $\Psi(E)$ are those that 
correspond to the ones in the stationary reservoir 
with the same $\psi_E$.

Then we construct operators $\hat a_{\alpha}(E)/\hat a^{\dagger}_{\alpha}(E)$
corresponding to incoming particles in 
the leads with definite energy $E$ as follows:
\begin{equation}
\label{smf_12}
\hat a_{\alpha}(E) =
\sum\limits_{n=-\infty}^{\infty}J_{n}\left(\frac{eV_{\alpha}}{\hbar\Omega_0}
\right) 
e^{-in\varphi_{\alpha}} \hat a^{\prime}_{\alpha}(E-n\hbar\Omega_0).
\end{equation} 
The operators $\hat b$ for scattered particles are related to the operators
$\hat a$ for incoming particles through the 
scattering matrix of the sample, Eq.(\ref{smf_3}).
Substituting the calculated $\hat a$ and $\hat b$ operators into Eq.(\ref{smf_1})
and averaging over the time-period we get a dc current in the presence of oscillating potentials 
$V_{\alpha}(t)$ at reservoirs as follows:
\begin{equation}
\label{smf_13}
\begin{array}{c}
I_{\alpha} = \frac{e}{\hbar}\int\limits_{0}^{\infty}dE
\sum\limits_{n=-\infty}^{\infty} \bigg\{
\sum\limits_{\beta=1}^{N_r} f_{\beta}(E - n\hbar\Omega_0) \\
\ \\
\times \sum\limits_{m,q=-\infty}^{\infty}
S^{*}_{F,\alpha\beta}(E,E_q) S_{F,\alpha\beta}(E,E_m) \\
\ \\
\times
J_{n+q}\left(\frac{eV_{\beta}}{\hbar\Omega_0} \right)
J_{n+m}\left(\frac{eV_{\beta}}{\hbar\Omega_0} \right) e^{i(q-m)\varphi_{\beta}} \\
\ \\
- f_{\alpha}(E - n\hbar\Omega_0)
J^{2}_{n}\left(\frac{eV_{\alpha}}{\hbar\Omega_0} \right) \bigg\}.
\end{array}
\end{equation}
Using Eq.(\ref{smf_4B}) and making the shift $E\to E + n\hbar\Omega_0$ in the
 last term one can rewrite the above equation in a
 more convenient form:
\begin{equation}
\label{smf_14}
\begin{array}{c}
I_{\alpha} = \frac{e}{\hbar}\int\limits_{0}^{\infty}dE
\sum\limits_{n=-\infty}^{\infty} \sum\limits_{\beta=1}^{N_r} 
\Big\{f_{\beta}(E - n\hbar\Omega_0) - f_{\alpha}(E) \Big\} \\
\ \\
\times \sum\limits_{m,q=-\infty}^{\infty}
S^{*}_{F,\alpha\beta}(E,E_q) S_{F,\alpha\beta}(E,E_m) \\
\ \\
\times
J_{n+q}\left(\frac{eV_{\beta}}{\hbar\Omega_0} \right)
J_{n+m}\left(\frac{eV_{\beta}}{\hbar\Omega_0} \right) e^{i(q-m)\varphi_{\beta}}. 
\end{array}
\end{equation}

To calculate the current flowing through the driven mesoscopic sample it is
necessary to know the Floquet 
scattering matrix $\hat S_{F}$. 
At strong driving this matrix has an infinite number of elements.
Therefore, its calculation is, in practice, a non-trivial problem.
This problem can be greatly simplified if the driving frequency is small.
In this limit the Floquet scattering matrix $\hat S_{F}$ can be related to the
stationary scattering matrix $\hat S_{0}$ having 
much less number of elements.

\subsection{Adiabatic approximation}

At low driving frequencies, $\Omega_0\to 0$, one can expand the elements of the Floquet scattering matrix in powers of $\Omega_0$. 
The lowest-order terms read as follows: \cite{micha3}
\begin{subequations}
\label{smf_15}
\begin{equation}
\label{smf_15A}
\begin{array}{c}
\hat S_{F}(E_n,E)  = \hat S_{0,n}(E) + \frac{n\hbar\Omega_0}{2}
\frac{\partial \hat S_{0,n}(E)}{\partial E} \\
\ \\
 + \hbar\Omega_0\hat A_{n}(E) + {\cal O}(\Omega_0^2),
\end{array}
\end{equation}
\begin{equation}
\label{smf_15B}
\begin{array}{c}
\hat S_{F}(E,E_n)  = \hat S_{0,-n}(E) + \frac{n\hbar\Omega_0}{2}
\frac{\partial \hat S_{0,-n}(E)}{\partial E} \\
\ \\
+ \hbar\Omega_0\hat A_{-n}(E) + {\cal O}(\Omega_0^2).
\end{array}
\end{equation}
\end{subequations}
Here $\hat S_{0,n}$ is the Fourier transform for the frozen scattering matrix which is defined as follows:
\begin{equation}
\label{smf_16}
\hat S_{0}(E,t) = \sum\limits_{n=-\infty}^{\infty} e^{-in\Omega_0 t}\hat S_{0,n}(E).
\end{equation} 
 Let the stationary scattering matrix $\hat S_0(E)$ depend on some parameters 
$p_i \in \{P\}, i=1, 2, \dots, N_p$ which are varied under an external
drive. Since we assume that the latter is periodic, 
the parameters are periodic in time as well,  $p_i(t+\tau_0) = p_i(t)$.
The frozen scattering matrix $\hat S_{0}(E,t)$ is defined as the stationary scattering matrix 
$\hat S_{0}(E,\{P\})$ with parameters being dependent on time:
$\hat S_{0}(E,t) = \hat S_{0}(E,\{P(t)\})$.
We emphasize that the frozen scattering matrix does not define the scattering
properties of a driven scatterer. Only the Floquet 
scattering matrix does.

The matrix $\hat A(E,t)$, whose Fourier elements $\hat A_{n}(E)$ that enters
the expansion (\ref{smf_15}), 
cannot be related to the stationary scattering matrix and have to be
calculated independently, see Ref.\onlinecite{micha3} for 
simple examples.
Notice that the current conservation introduces some constraints to the matrix $\hat A$.
Substituting Eq.(\ref{smf_15}) into Eq.(\ref{smf_4}) and taking into account
that the stationary scattering matrix is unitary we 
get the following: \cite{micha2}
\begin{equation}
\label{smf_17}
\hbar\Omega_0\Big\{ \hat S_{0}^{\dagger}\hat A +
\hat A^{\dagger}\hat S_{0}\Big\} = \frac{i\hbar}{2}\left( 
\frac{\partial\hat S_{0}^{\dagger}}{\partial t}
\frac{\partial\hat S_{0}}{\partial E} - 
\frac{\partial\hat S_{0}^{\dagger}}{\partial E}
\frac{\partial\hat S_{0}}{\partial t}\right).
\end{equation}
The matrix $\hat A$ reflects a directional asymmetry of a dynamical scattering
process arising as a result of interference of 
photon-assisted scattering amplitudes.
\cite{BM06}

Equations (\ref{smf_15}) and (\ref{smf_17}) show that the expansion in powers
of $\Omega_0$ is, in fact, an expansion in powers of 
$\hbar\Omega_0/\delta E$, where $\delta E$ is an energy scale characteristic for the stationary scattering matrix. 
The energy scale $\delta E$ relates to the inverse time spent by an electron inside the scattering region (the dwell time).
Therefore, alternatively one can say that the adiabatic expansion given in
Eq.(\ref{smf_15}), can be applied if the period of an 
external drive is large compared with the dwell time.
This definition of an adiabatic regime is different from a usually used in
quantum mechanics one which relates the excitation 
quantum $\hbar\Omega_0$ to the electron level spacing.

We conclude, to find the Floquet scattering matrix with accuracy of order
$\Omega_0$ it is necessary to calculate two matrices, 
$\hat S_{0}(E,t)$ and $\hat A(E,t)$, each of them having 
$N_r\times N_r$ elements.  If this is done we can calculate the current up to terms of order $\Omega_0$. 
For the simplicity we suppose that all the reservoirs have the same chemical
potentials and temperatures, 
hence $f_{\alpha}(\omega) = f_0(\omega), \forall \alpha$. 
Then substituting Eq.(\ref{smf_15}) into Eq.(\ref{smf_14}), expanding the
difference of Fermi functions in powers of 
$\Omega_0$, performing inverse Fourier transformation, and keeping at the end
only linear in $\Omega_0$ and $V_{\alpha}(t)$ terms we get 
the dc current as a sum of three contributions:
\begin{subequations}
\label{smf_18}
\begin{eqnarray}
\label{smf_18A}
I_{\alpha} &  = & 
\int\limits_{0}^{\infty}dE 
\left(-\frac{\partial f_{0}(E)}{\partial E} \right)
\int\limits_{0}^{\tau_0} \frac{dt}{\tau}\Big\{
I_{\alpha}^{(pump)}(E,t) + \nonumber \\
& & I_{\alpha}^{(rect)}(E,t) + I_{\alpha}^{(int)}(E,t) \Big\},  
\end{eqnarray}
\begin{equation}
\label{smf_18B}
I_{\alpha}^{(pump)}(E,t) = -i\frac{e}{2\pi}
\left(\hat S_{0}(E,t)\frac{\partial\hat S_{0}^{\dagger}(E,t)}{\partial t} \right)_{\alpha\alpha},
\end{equation}
\begin{equation}
\label{smf_18C}
I_{\alpha}^{(rect)}(E,t) = \frac{e^2}{h}\sum\limits_{\beta=1}^{N_r} 
\Big\{V_{\beta}(t) - V_{\alpha}(t) \Big\}
\Big| S_{0,\alpha\beta}(E,t)\Big|^2,
\end{equation}
\begin{equation}
\label{smf_18D}
\begin{array}{c}
I_{\alpha}^{(int)}(E,t) = \frac{e^2}{h} \sum\limits_{\beta=1}^{N_r} 
V_{\beta}(t) \bigg\{2\hbar\Omega_0{\rm Re}
\big[S_{0,\alpha\beta}^{*}(E,t)A_{\alpha\beta}(E,t) \big] \\
\ \\
+ \frac{i\hbar}{2}\left( 
\frac{\partial S_{0,\alpha\beta}}{\partial t}
\frac{\partial S_{0,\alpha\beta}^{*}}{\partial E} -
\frac{\partial S_{0,\alpha\beta}}{\partial E} 
\frac{\partial S_{0,\alpha\beta}^{*}}{\partial t}
\right) \bigg\}.
\end{array}
\end{equation}
\end{subequations}
Here, in Eq.(\ref{smf_18D}), we have integrated over energy the term
proportional to $\partial^2f_{0}/\partial E^2$ and made 
it to be proportional to $\partial f_{0}/\partial E$.

The first contribution, $I_{\alpha}^{(pump)}(E,t)$, is an adiabatic current pumped by a dynamical scatterer. \cite{Brouwer98}
This current is non zero if the time-reversal symmetry in the system is broken by the external drive, $\hat S_{0}(t)\ne\hat S_{0}(-t)$. 
To this end at least two parameters of the scatterer have to be varied with a phase lag different from zero and $\pi$. 
The second contribution, $I_{\alpha}^{(rect)}(E,t)$, is due to the rectification of
ac currents flowing under the influence of ac
 voltages by the varying conductance of a sample.
\cite {Brouwer01}
This contribution is non-zero if the potentials $V_{\alpha}(t)$ are different.
The third contribution,  $I_{\alpha}^{(int)}(E,t)$, is due to interference between
the ac currents generated by the dynamical 
scatterer and the ac currents produced by the external voltages.
\cite{micha2} This current can be non-zero even if all the potentials
$V_{\alpha}(t)$ are the same. Formally, this contribution 
can be viewed as due to external voltages acting as additional pumping parameters. 
So, if all the potentials are the same, $V_{\alpha}(t)=V(t), \forall \alpha$
and only one parameter of a scatterer is varied then 
only the third contribution remains.

Now, we turn to the Keldysh formalism to quantum transport for harmonically driven
systems and then, we will establish a correspondence 
between the two formalisms.

\section{Quantum transport within Keldysh formalism}
\subsection{Hamiltonian, Green's functions and Dyson equations}
The goal of Keldysh formalism is the evaluation of the Green's function 
of the system in the real time axis. The starting point is a Hamiltonian
 to describe the driven system in contact
to the $N_r$ particle reservoirs through connecting leads: 
\begin{equation}
H(t)=H^{sys}(t) + H^{cont} +  H^{res}(t). 
\end{equation}
Although this formalism provides a systematic way to approximately
treat many-body interactions,  we focus in  Hamiltonians
corresponding to non-interacting electrons, which can be treated exactly.  
In several problems, it is convenient to describe the central driven system by
a lattice model containing  $N$ sites. We adopt here that point of view.
 For such a system a generic
Hamiltonian of spinless noninteracting electrons  reads
\begin{equation}
H^{sys}(t) = \sum_{l,l'=1}^N (H^{sys}_{l,l'}(t) c^{\dagger}_l c_{l'} + H.c.).
\label{sys}
\end{equation}
We assume that the matrix elements contain
static and time-dependent pieces of the form:
$H^{sys}_{l,l'}(t)= \varepsilon_{l,l'}(\Phi)+ V_{l,l'}(t,\delta_{l,l'})$,
being $\Phi$ a static magnetic flux and the terms depending on the  
driving fields being periodic in time,
$V_{l,l'}(t,\delta_{l,l'})= 
\sum_{k= - \infty }^{\infty} e^{-i k \Omega_0 t} V_{l,l'} (k) $
with amplitudes depending on the phases $\delta_{l,l'}$ and $ V_{l,l'}(0)=0$. For the moment,
we do not write explicitly the dependence of the matrix elements of $H^{sys}$ on the magnetic flux and 
the phases. Instead, we 
simply write $\varepsilon_{l,l'}$ and $V_{l,l'}(t)$. We shall recover
the more complete notation in section VI, where we shall analyze symmetry
properties that depend on those parameters.  In order to simplify the notation
 we also adopt energy units with $\hbar=1$.

 The contacts between the system and the reservoirs are described
by hopping terms of the form 
\begin{equation}
H^{cont}= - \sum_{\alpha  } w_{\alpha}
(c^{\dagger}_{k_{\alpha}} c_{j_{\alpha}} + H.c.),
\label{cont}
\end{equation}
 where $k_{\alpha}$ 
and $j_{\alpha}$ are, respectively, coordinates of the reservoirs and the 
central system.
We assume models of free electrons for the reservoirs,
\begin{equation}
H^{res}=  \sum_{\alpha k_{\alpha} } \varepsilon_{k_{\alpha}}
c^{\dagger}_{k_{\alpha}} c_{k_{\alpha}}.
\label{res}
\end{equation}
We  also consider 
the possibility of applying ac-external voltages at the reservoirs, which are represented by
potentials of the form (\ref{smf_10}). This introduces 
a time-dependent shift in the energies $\varepsilon_{k_{\alpha}} \rightarrow
\varepsilon_{\alpha}(t)=\varepsilon_{k_{\alpha}} + 
e V_{\alpha} \cos(\Omega_0 t +\varphi_{\alpha})$.
Alternatively, it is possible to get rid of these time-dependent
shifts by recourse to a gauge transformation
$c_{k_{\alpha}}  \rightarrow  \overline{c}_{k_{\alpha}} (t)=
c_{k_{\alpha}} e^{-i \int_{t_0}^t dt_1  
e V_{\alpha} \cos(\Omega_0 t_1 +\varphi_{\alpha})}$.
Within such a procedure, the hopping 
matrix element of (\ref{cont}) becomes modified to 
$w_{\alpha} \rightarrow w_{\alpha} e^{-i \int_{t_0}^t dt_1  
e V_{\alpha} \cos(\Omega_0 t_1 +\varphi_{\alpha})}$. Both ways of tackling the
problem end in the same results for the evaluated mean values of observables.
We choose the first one.

The elementary theoretical tools of Keldysh formalism are the retarded, lesser, 
advanced, and bigger  Green's functions. The first ones are defined as follows 
\begin{subequations}
\begin{equation}
G^R_{l,l'}(t,t')   =    \Theta (t-t') \Big[ G_{l,l'}^>(t,t') -
G_{l,l'}^<(t,t') \Big],
\label{gret}
\end{equation}
\begin{equation}
G_{l,l'}^<(t,t')  =   i \langle  c^{\dagger}_{l'}(t')  c_l(t) \rangle,
\label{gles}
\end{equation}
\end{subequations}
while the remaining functions are defined from the relations:
 $G_{l,l'}^>(t,t')=  [G_{l',l}^<(t',t)]^*$ and
$G_{l,l'}^A(t,t')= [G_{l',l}^R(t',t)]^*$.
These Green's functions are used in the calculation of mean values of observables, in
 particular, the currents flowing through the different pieces of the driven
 system. Their evaluation needs the solution of the equations
 governing their time-evolution, which are derived from a matricial Dyson
 equation \cite{kel}.  If we focus on spatial coordinates lying on the 
lattice of the central system it is convenient to define the matrices
 $ \hat{H}^{sys}(t)$ and $\hat{G}^{R,<}(t, t')$, which have
matrix elements $ H^{sys}_{l,l'}(t)$
and $G^{R,<}_{l,l'}(t, t')$, respectively. The corresponding matrices for
the advanced and the bigger Green's functions are obtained from the relations
$\hat{G}^{A}(t,t') = [\hat{G}^{R}(t',t)]^{\dagger}$ and
$\hat{G}^{>}(t,t') = [\hat{G}^{<}(t',t)]^{\dagger}$.
The Dyson equations for these Green's functions are:
\begin{subequations}
\begin{equation}
\begin{array}{c}
 \Big\{ i \frac{\overrightarrow{\partial}}{\partial t} - \hat{H}^{sys}(t) \Big\} \hat{G}^R(t, t')
 \\ \ \\
- \int_{t'}^t dt_1
\hat{\Sigma}^R(t,t_1) \hat{G}^R(t_1, t') = \hat{1} \delta(t-t'),  \\ \ \\
 \hat{G}^R(t, t') \Big\{-i \frac{\overleftarrow{\partial}}{\partial t'}  
-  \hat{H}^{sys}(t') \Big\}
 \\ \ \\
- \int_{t'}^t dt_1
\hat{G}^R (t, t_1) \hat{\Sigma}^R (t_1,t') = \hat{1}  \delta(t-t'),
\label{difdyr}
\end{array}
\end{equation}
\begin{equation}
\begin{array}{c}
\Big\{ i \frac{\overrightarrow{\partial}}{\partial t}  - \hat{H}^{sys}(t)
\Big\}\hat{G}^<(t, t')
- \int_{-\infty}^t dt_1
\hat{\Sigma}^R(t,t_1)\hat{G}^<(t_1, t') \\ \ \\
- \int_{-\infty}^{t'} dt_1 \hat{\Sigma}^<(t,t_1) \hat{G}^A(t_1, t')   
=\hat{ 0}, \\ \ \\
 \hat{ G}^<(t, t') \Big\{ -i \frac{\overleftarrow{\partial}}{\partial t'}-
 \hat{H}^{sys}(t') \Big\}-   \int_{-\infty}^t dt_1
\hat{G}^R(t, t_1)  \hat{\Sigma}^<(t_1,t') \\ \ \\
- \int_{-\infty}^{t'} dt_1\hat{G}^<(t, t_1) \hat{\Sigma}^A(t_1,t')    = \hat{0},
\label{difdyl}
\end{array}
\end{equation}
\end{subequations}
where $\overrightarrow{\partial}$ and $\overleftarrow{\partial}$ indicates
that the operator acts to the right and to the left, respectively.
The Dyson equations for the Green's functions with spatial coordinates
outside the central region have a similar structure, considering the
corresponding pieces of the Hamiltonian $H$ instead of $\hat{H}^{sys}(t)$ and $\hat{\Sigma}^{R, >}(t,t')=0$.
 In the absence of many-body interactions, the  self-energies entering
 (\ref{difdyr}) and (\ref{difdyl}) take into
account only
the
effect of the ``escape to the leads''. 
They
are obtained from the Dyson equations  for Green's functions with coordinates along
the connections to the leads and integrating out the degrees of freedom related
to the reservoirs. \cite{past,JWM94} Explicitly, they read:
\begin{equation}
\Sigma^{^{<}_{ >}}_{l,l'}(t, t') = \delta_{l, j_{\alpha}}
\delta_{l', j_{\alpha}} |w_{\alpha}|^2 
g^{^{<}_{ >}}_{\alpha}(t, t'),
\end{equation}
 and 
\begin{equation}
\Sigma^R_{l,l'}(t,t')= \Theta(t-t') \Big[
\Sigma^>_{l,l'}(t,t')-\Sigma^<_{l,l'}(t,t') \Big],
\end{equation}
 with
\begin{eqnarray}
& & g^{^{<}_{ >}}_{\alpha}(t, t') = 
\pm i \sum_{k,m} e^{-i k \Omega_0 t}
J_m(\frac{V_{\alpha}}{\Omega_0})J_{m+k}(\frac{V_{\alpha}}{\Omega_0}) 
e^{-i k \varphi_{\alpha} }
 \nonumber \\
& & \times  \int_{-\infty}^{\infty} \frac{d \omega}{2 \pi} e^{-i \omega (t-t')} 
\lambda^{^{<}_{>}}_{\alpha}(\omega- m \Omega_0) \rho_{\alpha}(\omega- m \Omega_0) , 
\label{self}
\end{eqnarray} 
where the density of states $\rho_{\alpha}(\omega)= \sum_{k_{\alpha}}
\delta(\omega-\varepsilon_{k_{\alpha}} )$, corresponds to  $H^{res}$, while
 $ \lambda^{<}_{\alpha}(\omega)= f_{\alpha}(\omega)$ and 
$ \lambda^{>}_{\alpha}(\omega)= 1- f_{\alpha}(\omega)$, being
$f_{\alpha}(\omega)=1/(e^{\beta_{\alpha}(\omega - \mu^0_{\alpha})}+1)$,
the Fermi function corresponding to the reservoir $\alpha$, which is assumed to
be at the
temperature $1/\beta_{\alpha}$.  This function is
only well defined for equilibrium problems and is thus  related  only to the
stationary component of the Hamiltonian $H^{res}$. \cite{JWM94}  
We also define the functions
\begin{eqnarray}
\Gamma_{\alpha}(k,\omega)& = & |w_{\alpha}|^2 e^{-i k \varphi_{\alpha} }
\sum_{m=-\infty}^{\infty}
J_m(\frac{V_{\alpha}}{\Omega_0})J_{m+k}(\frac{V_{\alpha}}{\Omega_0}) \nonumber \\
& & \times \Gamma^0_{\alpha}(\omega- m \Omega_0), \nonumber \\
\Gamma^{^{<}_{>}}_{\alpha}(k,\omega)& = & |w_{\alpha}|^2 e^{-i k \varphi_{\alpha} }
\sum_{m=-\infty}^{\infty}
J_m(\frac{V_{\alpha}}{\Omega_0})J_{m+k}(\frac{V_{\alpha}}{\Omega_0}) \nonumber
\\
& & \times
\lambda^{^{<}_{>}}_{\alpha}(\omega-m \Omega_0) \Gamma^0_{\alpha}(\omega- m \Omega_0), \nonumber \\
\Gamma^0_{\alpha}(\omega)& = &|w_{\alpha}|^2 \rho_{\alpha}(\omega).
\label{gammas}
\end{eqnarray}

In the practical solution of the problem, the strategy followed in Ref. \onlinecite{lilip,lilir}
was to work  with convenient  integral representations of  the Dyson equations
(\ref{difdyr}) and (\ref{difdyl}):
\begin{subequations}
\begin{eqnarray}
& & \hat{G}^R(t,t') = \hat{G}^0(t-t') + \nonumber \\
& &  {\sum_{k=-\infty}^{\infty}}^{\prime} \int_{t'}^t dt_1 e^{-i k \Omega_0 t_1} 
\hat{G}^R(t,t_1)\hat{V}(k) \hat{G}^0(t_1-t')
+ \nonumber \\
& & {\sum_{k=-\infty}^{\infty}}^{\prime}
\int_{t'}^t dt_1 \int_{t'}^t dt_2 \frac{d\omega}{2\pi} e^{-i k \Omega_0 t_1} e^{-i \omega (t_1-t_2)}
\nonumber \\
& &
\times
\hat{G}^R(t,t_1) \hat{\Sigma}(k,\omega) \hat{G}^0(t_1-t'),
\label{dyre}
\end{eqnarray}
\begin{eqnarray}
& & \hat{G}^{^{<}_{>}}(t,t')  =  
\int_{-\infty}^t dt_1 \int_{-\infty}^{t'} dt_2 \hat{G}^R(t,t_1)
\hat{\Sigma}^{^{<}_{>}}(t_1 , t_2) \nonumber \\
& & \times
\hat{G}^A(t_2,t'),
\label{dyeq}
\end{eqnarray}
\end{subequations}
where ${\sum_k}^{\prime}$ denotes $\sum_{k\neq 0}$.
For the coordinates $(j_{\alpha} , k_{\alpha}) $ along the contact, it is convenient to work with
\begin{subequations}
\begin{equation}
\begin{array}{r}
 G^R_{j_{\alpha},\alpha}(t,t')=
-w_{\alpha} \int_{-\infty}^t dt_1  G^R_{j_{\alpha},j_{\alpha}}(t,t_1)
g^R_{\alpha}(t_1, t'), \label{dycontr}
\end{array}
\end{equation}
\begin{equation}
\begin{array}{l}
 G^<_{j_{\alpha},\alpha}(t,t')=
-w_{\alpha} \Big\{ \int_{-\infty}^t dt_1  G^R_{j_{\alpha},j_{\alpha}}(t,t_1)
g^<_{\alpha}(t_1, t') \\ \ \\
 +
 \int_{-\infty}^{t'} dt_1 G^<_{j_{\alpha},j_{\alpha}}(t,t_1)
g^A_{\alpha}(t_1, t') \Big\},
\label{dycont}
\end{array}
\end{equation}
\end{subequations}
where we have defined $ G^{R,<}_{j_{\alpha},\alpha}(t,t')= \sum_{k_{\alpha}}
  G^{R,<}_{j_{\alpha},k_{\alpha}}(t,t')$, while $g^R_{\alpha}(t, t')=
  \Theta(t-t') [g^>_{\alpha}(t, t') - g^<_{\alpha}(t, t')]$ and $g^A_{\alpha}(t,
  t')=[g^R_{\alpha}(t', t)]^*$.
In equations (\ref{dyre}) and (\ref{dyeq}), we have also defined:
\begin{equation}
\hat{\Sigma} (k,\omega)  =  \int_{-\infty}^{\infty} \frac{d \omega'}{2 \pi}
\frac{\hat{\Gamma} (k, \omega' )}{\omega - \omega'+ i 0^+},
\label{sigma}
\end{equation}
where $\hat{\Gamma} (k, \omega)$ has matrix elements
${\Gamma}_{l,l'} (k, \omega)= \delta_{l,j_{\alpha}} \delta_{l',j_{\alpha}}
{\Gamma}_{\alpha} (k, \omega ) $.
The retarded Green's function $\hat{G}^0(t-t')$
corresponds to the equilibrium problem defined by the static piece of $H^{sys}(t)$
(with matrix elements $\varepsilon_{l,l'}$) dressed by the
static component of the self-energy. In other words, it is the solution of
\begin{equation}
[\omega \hat{1} - \hat{\varepsilon}- \hat{\Sigma}(0,\omega)] \hat{G}^0(\omega)=\hat{1},
\label{dy0}
\end{equation}
being
\begin{equation}
\hat{G}^0(\omega)=\int_{-\infty}^t dt'\hat{G}^0(t-t') 
e^{i (\omega + i 0^+)(t-t')}, \label{fou0}
\end{equation}

Notice that in the equation for the lesser (bigger) Green's function,
 we have dropped a term that depends on the solution of the
homogeneous equation, which is only relevant in the description of transient behavior.

According to Refs.~\onlinecite{lilip,lilir}, we introduce the following Fourier transform
for the retarded Green's function:
\begin{equation}
\hat{G}^R(t,\omega)= \int_{-\infty}^t dt' \hat{G}^R(t,t') 
e^{i (\omega + i 0^+)(t-t')}.
\label{fourfre}
\end{equation}
Transforming (\ref{dyre}) according to (\ref{fourfre}) results in the following
 set of linear equations
\begin{eqnarray}
& & \hat{G}^R(t,\omega) = \hat{G}^0(\omega) + \nonumber \\
& &    {\sum_{k=-\infty}^{\infty}}^{\prime} e^{-i k \Omega_0 t}
 \hat{G}^R(t,\omega+k\Omega_0) \hat{V}(k) \hat{G}^0(\omega) + \nonumber \\
& &   {\sum_{k=-\infty}^{\infty}}^{\prime}   e^{-i k \Omega_0 t}
\hat{G}^R(t,\omega+k\Omega_0) \hat{\Sigma} (k, \omega )
\hat{G}^0(\omega).
\label{retfre}
\end{eqnarray} 
Since the above equation is periodic in $t$ with period $\tau_0=\Omega_0/2\pi$, it is possible
to expand its solution in Fourier series:
\begin{equation}
\hat{G}^R(t,\omega)= \sum_{k=-\infty}^{\infty} e^{-ik \Omega_0 t} 
\hat{\cal G} (k,\omega).
\label{disfre}
\end{equation}
Note that the above procedure takes care of causality. In fact,
the transformation (\ref{fourfre}), defined with respect to the difference of
the two times is the natural extension to the transformation (\ref{fou0})
for retarded Green's functions defined in text books for stationary problems, which 
ensures correct analytical properties of the transformed function. 
In addition, notice that  (\ref{disfre}) is a Fourier series,  not a Fourier transformation,
which reflects the fact that Dyson equation is periodic in the ``observational time''
$t$ and so does the corresponding solution. 
The retarded Green's function can be calculated from the solution of
the linear set (\ref{retfre}).  A convenient method for the direct evaluation
of the Fourier components (\ref{disfre}) is the renormalization method of Ref.~\onlinecite{lilire}.

\subsection{Current through the leads}
The  mean value of observables related to one-body operators
can be directly expressed in terms of the lesser (or bigger) Green's function.
In particular, the charge current flowing through the lead from the central system
towards the
reservoir
$\alpha$,   can be written (in units of $e/\hbar$) as:
\begin{eqnarray}
J_{\alpha}(t) & = & i w_{\alpha} \sum_{k_{\alpha}} \langle
  c^{\dagger}_{k_{\alpha}}  c_{j_{\alpha}} -  c^{\dagger}_{j_{\alpha}}
  c_{k_{\alpha}} \rangle \nonumber \\
& = &  2 w_{\alpha} \mbox{Re} \Big[ G^<_{j_{\alpha},\alpha}(t,t) \Big].
\label{cur0}
\end{eqnarray}
Notice that the above current is related to the current (\ref{smf_1}) through:
$I_{\alpha}(t)= e J_{\alpha}(t)/\hbar $.
Taking into account the Dyson equation describing the contact 
between the central system  and the lead (\ref{dycont})
this current can be written as:
\begin{eqnarray}
& & J_{\alpha}(t) = -2 |w_{\alpha}|^2 \mbox{Re} \Big\{  \int_{-\infty}^t dt_1 
\Big[ G^R_{j_{\alpha},j_{\alpha}}(t,t_1) g^<_{\alpha}(t_1, t) \nonumber \\
& &
+
G^<_{j_{\alpha},j_{\alpha}}(t,t_1) g^A_{\alpha}(t_1, t) \Big] \Big\} .
\label{1}
\end{eqnarray}

Making use of the definition (\ref{gret}),  
we can also express (\ref{1}) as follows:
\begin{eqnarray}
& & J_{\alpha}(t)  = - 2 |w_{\alpha}|^2 \mbox{Re} \Big\{ \int_{-\infty}^t
d t_1 \Big[G^>_{j_{\alpha}, j_{\alpha}}(t,t_1)-
\nonumber \\
& &
G^<_{j_{\alpha}, j_{\alpha}}(t,t_1) \Big]
g^<_{\alpha}(t_1,t)
+ G^>_{j_{\alpha}, j_{\alpha}}(t,t_1) g^A_{\alpha}(t_1,t) \Big\},
\label{curtt}
\end{eqnarray}
which, for the case of stationary reservoirs, simplifies to:
\begin{equation}
\begin{array}{l}
J_{\alpha}(t)= 
 2 \mbox{Im} \Big\{  \int_{-\infty}^t dt_1 \int_{-\infty}^{+\infty} \frac{d \omega}{2 \pi} 
e^{i \omega (t-t_1)} \Gamma^0_{\alpha}(\omega) \times  \\ \ \\
\Big[ G^>_{j_{\alpha},j_{\alpha}}(t,t_1) f_{\alpha}(\omega) + 
G^<_{j_{\alpha},j_{\alpha}}(t,t_1) \Big( 1- f_{\alpha}(\omega) \Big) \Big] \Big\}.
\label{curtbol}
 \end{array}
\end{equation}
The above expression has an appealing form, since it depends on
Boltzmann-like factors. In fact, $ \Gamma^0_{\alpha}(\omega) f_{\alpha}(\omega)
G^>_{j_{\alpha},j_{\alpha}}(t,t_1)$ is related to the probability for a state
in the  
lead $\alpha$ to
be occupied times the probability for its closest site at the central structure,
$j_{\alpha}$,
to be unoccupied, while the term 
$( 1- f_{\alpha}(\omega))\Gamma^0_{\alpha}(\omega)G^<_{j_{\alpha},j_{\alpha}}(t,t_1) $ 
 is related to the probability of the opposite 
process to take place.

In what follows, we focus in the
dc-component of the current, which is defined as:
\begin{equation}
J^{dc}_{\alpha}= \frac{1}{\tau_0} \int_0^{\tau_0} dt J_{\alpha}(t).
\label{dc}
\end{equation}

\subsubsection{Review of the stationary case}
Let us first consider $V_{l,l'}(t)= V_{\alpha}(t)=0$ and we 
review the procedure introduced by Caroli et al. \cite{caro}
Since in the stationary regime
the Green's functions entering eq. (\ref{1}) depend only on the
difference of times, it is possible to perform the usual Fourier transform
in the variable $t-t'$. The result is:
\begin{eqnarray}
J_{\alpha}& = & - 2 |w_{\alpha}|^2 \mbox{Re} \Big\{ \int_{-\infty}^{\infty} \frac{d\omega}{2\pi}
  \Big[ G^R_{j_{\alpha}, j_{\alpha} }(\omega) g^<_{\alpha}(\omega) \nonumber \\
& & +
G^<_{j_{\alpha}, j_{\alpha}}(\omega) g^A_{\alpha}(\omega) \Big] \Big\}.
\label{2}
\end{eqnarray}
Then, using the following identities and definitions:
\begin{eqnarray}
& & g^R_{\alpha}(\omega)=[g^A_{\alpha}(\omega)]^*= \int_{-\infty}^{\infty}
\frac{d \omega'}{2 \pi} \frac{ \rho_{\alpha}(\omega')}
{ \omega - \omega'+ i 0^+}, \nonumber\\
& & g^<_{\alpha}(\omega)= i f_{\alpha}(\omega) \rho_{\alpha}(\omega), 
\end{eqnarray}
as well as the Dyson equation:
\begin{equation}
G^<_{j_{\alpha} j_{\alpha}}(\omega)=
\sum_{\beta=1}^{N_r}  G^R_{j_{\alpha}, j_{\beta}}(\omega)
\Sigma^<_{j_{\beta}}(\omega)G^A_{j_{\beta}, j_{\alpha}}(\omega),
\end{equation}
Eq. (\ref{2}) can be written as:
\begin{eqnarray}
J_{\alpha} & = &  |w_{\alpha}|^2 \int_{-\infty}^{\infty}
 \frac{d\omega}{2\pi}
 \Big\{
f_{\alpha}(\omega) \rho_{\alpha}(\omega) 
2 \mbox{Im} \Big[ G^R_{j_{\alpha}, j_{\alpha}}(\omega) \Big]\nonumber\\
& & + \sum_{\beta} f_{\beta}(\omega) \Gamma^0_{\beta}(\omega)
|G^R_{j_{\alpha}, j_{\beta}}(\omega)|^2 \Big\}.
\label{eqn}
\end{eqnarray}
Let us note that in the present case $\hat{G}^R(\omega)\equiv
\hat{G}^0(\omega)$ with $\hat{\Sigma}(0,\omega) \equiv
\hat{\Sigma}^0(\omega)$,
being
\begin{equation} 
{\Sigma}_{l,l'}^0(\omega)=\sum_{\alpha=1}^{N_r} \delta_{l,j_{\alpha}}
\delta_{l',j_{\alpha}} \int \frac{d\omega'}{2 \pi}
\frac{\Gamma^0_{\alpha}(\omega')}{\omega-\omega'+ i 0^+}.
\end{equation}
Being an  equilibrium Green's function, $\hat{G}^0(\omega)$
 satisfies the following  property (see appendix A for a proof):
\begin{equation} 
 \mbox{Im}\Big[ G^0_{j_{\alpha}, j_{\alpha}}(\omega) \Big] = 
\sum_{\beta=1}^{N_r}
G^0_{j_{\alpha}, j_{\beta}}(\omega)   \Gamma^0_{\beta} (\omega)  
G^0_{j_{\alpha}, j_{\beta}}(\omega)^*.
\label{cruc}
\end{equation}
Using it in (\ref{eqn}) and recalling the definition of the
self-energy (\ref{self}),
we get the well known expression for the current \cite{caro}:
\begin{equation}
J_{\alpha}=\sum_{\beta=1}^{N_r} \int_{-\infty}^{\infty} \frac{d\omega}{2\pi}  
\Gamma^0_{\alpha}(\omega) \Gamma^0_{\beta}(\omega) 
|G^0_{j_{\alpha}, j_{\beta}}(\omega)|^2
\Big[ f_{\beta}(\omega)-f_{\alpha}(\omega) \Big].
\label{curest}
\end{equation}

\subsubsection{Time-periodic local potentials  with stationary reservoirs.}
Let us now consider the possibility of time-dependent
terms in the Hamiltonian of the central system, but stationary reservoirs,
i.e $V_{\alpha}(t)=0$.

Using (\ref{dyeq}) as well as the Fourier representation (\ref{disfre})
in Eq. (\ref{1}) we get the following expression for the dc-component
of the current through the lead $\alpha$:
\begin{equation}
\begin{array}{r}
J^{dc}_{\alpha} =  
\int_{-\infty}^{\infty} \frac{d\omega}{2\pi} \Big\{
2 \mbox{Im}\Big[{\cal G}_{j_{\alpha}, j_{\alpha}}(0,\omega) \Big] \Gamma^0_{\alpha}(\omega) 
f_{\alpha}(\omega) + \\
  \sum\limits_{\beta=1}^{N_r} \sum\limits_{k= -\infty}^{\infty}
| {\cal G}_{j_{\alpha}, j_{\beta}}(k,\omega) |^2  
f_{\beta}(\omega) \Gamma^0_{\beta}(\omega)
\Gamma^0_{\alpha}(\omega + k \Omega_0) \Big\}.
\label{jl}
\end{array}
\end{equation}

There are two additional equivalent expressions to the above dc-current, which
correspond to two different representations of the term
$\mbox{Im}\Big[{\cal G}_{j_{\alpha}, j_{\alpha}}(0,\omega) \Big]$.
The first one is obtained from the condition of the continuity of the current,
as shown in appendix B. An alternative proof is given in appendix C.
Substituting (\ref{idcont}) 
 in (\ref{jl}) results in the following representation:
\begin{eqnarray}
& & J^{dc}_{\alpha} =  \sum_{\beta \neq \alpha=1 }^{N_r}
\sum_{k=-\infty}^{\infty}  \int_{-\infty}^{\infty} 
\frac{d\omega}{2 \pi} \times \nonumber \\
& &
\Big\{  |{\cal G}_{j_{\alpha}, j_{\beta} }(k,\omega)|^2 \Gamma^0_{\beta }(\omega)  
\Gamma^0_{\alpha}(\omega + k \Omega_0)f_{\beta }(\omega) 
\nonumber \\
& &
- |{\cal G}_{j_{\beta}, j_{\alpha}}(k,\omega)|^2 \Gamma^0_{\alpha}(\omega)  
\Gamma^0_{\beta }(\omega + k \Omega_0)f_{\alpha}(\omega)
\Big\} .
\label{ref}
\end{eqnarray} 

The second additional representation corresponds to substituting  the
identity (\ref{shoap0}) derived in appendix D into (\ref{jl}). The result is:
\begin{eqnarray}
& & J^{dc}_{\alpha} = 
 \sum_{\beta =1 }^{N_r}
\sum_{k=-\infty}^{\infty} 
\int_{-\infty}^{\infty}  \frac{d\omega}{2\pi} 
\Gamma^0_{\beta }(\omega) \Gamma^0_{\alpha}(\omega+k \Omega_0) 
\nonumber \\
& &
\times |{\cal G}_{ j_{\alpha}, j_{\beta}}(k,\omega)|^2 
\Big[ f_{\beta }(\omega) - f_{\alpha}(\omega+k \Omega_0) \Big].
\label{curl}
\end{eqnarray} 

Any of the three representations (\ref{jl}), (\ref{ref}) and (\ref{curl}) are
equally valid to calculate the dc-current flowing through the lead
$\alpha$. 
The concrete evaluation implies the  solution of the retarded Green's function from
the  Dyson equation (\ref{retfre}) with $\hat{\Sigma}(k, \omega)=0$.

\subsubsection{Time-dependent voltages at the reservoirs.}
Let us finally consider the more general case, where, in addition to the
pumping potentials at the central structure, ac voltages are applied
at the reservoirs.

We start from the definition of the time-dependent current through lead
$\alpha$ (\ref{curtt}) and substitute there Eqs. (\ref{self}) and (\ref{dyeq}).
Then, we use the Fourier representation of the retarded Green's function
(\ref{disfre}) and 
take the dc-component. The result is: 
\begin{eqnarray}
& &J_{\alpha}^{dc}  =  \sum_{\beta =1 }^{N_r} \sum_{ k, q, p=-\infty}^{\infty} 
\mbox{Re} \Big\{ \int_{-\infty}^{\infty} \frac{d \omega}{2 \pi} \times
\nonumber \\
& &
[ \Gamma_{\alpha}(p, \omega + k \Omega_0) \Gamma^<_{\beta}(q, \omega )-
\Gamma^<_{\alpha}(p, \omega + k \Omega_0) \Gamma_{\beta}(q, \omega )]
\nonumber \\
& &
\times 
{\cal G}_{j_{\alpha}, j_{\beta} }(k-q+p,\omega+q \Omega_0)
{\cal G}^*_{j_{\alpha}, j_{\beta} }(k,\omega)  \Big\}.
\label{jacres0}
\end{eqnarray}
Equivalently, this expression can also be written as:
\begin{eqnarray}
& & J_{\alpha}^{dc}   =    \sum_{\beta =1 }^{N_r} \sum_{k, q, p=-\infty}^{\infty} 
J_{n+p}(\frac{e V_{\alpha}}{\Omega_0}) J_{n}(\frac{e V_{\alpha}}{\Omega_0})\nonumber \\
& & 
\times J_{m+q}(\frac{e V_{\beta}}{\Omega_0}) J_{m}(\frac{e V_{\beta}}{\Omega_0})
\mbox{Re} \Big\{ e^{i (p \varphi_{\alpha} + q \varphi_{\beta} )} 
\int_{-\infty}^{\infty} \frac{d \omega}{2 \pi} \Gamma^0_{\alpha}( \omega )
\nonumber \\
& & 
\times \Gamma^0_{\beta}( \omega + (n-k-m)\Omega_0 )
\Big[ f_{\beta}(\omega+ (n-k -m ) \Omega_0) 
\nonumber \\
& &
- f_{\alpha}(\omega) \Big] {\cal G}_{j_{\alpha}, j_{\beta} }(k-q+p,\omega+ (n-k+q) \Omega_0)\nonumber \\
& &
\times {\cal G}^*_{j_{\alpha}, j_{\beta} }(k,\omega +(n-k) \Omega_0) \Big\}.
\label{jacres}
\end{eqnarray}
 For reservoirs with a smooth density of states such that $\Gamma^0_{\alpha
 }(\omega -m \Omega_0)
 \sim \Gamma_{\alpha}(\omega)$ and 
$\Gamma^0_{\beta}(\omega-m \Omega_0) \sim \Gamma_{\beta}(\omega)$, the expression for the
dc-current further
simplifies. In fact, in such a case   (\ref{jacres0}) reduces to:
\begin{equation}
\begin{array}{l}
 J_{\alpha}^{dc}   =    \sum\limits_{\beta =1 }^{N_r} 
\sum\limits_{ k, q, m=-\infty}^{\infty} 
\mbox{Re} \Big\{  
\int_{-\infty}^{\infty} \frac{d \omega}{2 \pi} 
\Gamma^0_{\alpha}(\omega) \Gamma^0_{\beta}(\omega)
\Big[ e^{i q   \varphi_{\beta} } \times
 \\ \ \\
 f_{\beta}(\omega  -m  \Omega_0)
J_{m+q}(\frac{e V_{\beta}}{\Omega_0})
J_{m}(\frac{e V_{\beta}}{\Omega_0})
   {\cal G}_{j_{\alpha}, j_{\beta} }(k-q,\omega +q \Omega_0)
   \\ \ \\
{\cal G}^*_{j_{\alpha}, j_{\beta} }(k,\omega ) 
-  e^{i q   \varphi_{\alpha} } f_{\alpha}(\omega + (k-m) \Omega_0 )
J_{m+q}(\frac{e V_{\alpha}}{\Omega_0})  \\ \ \\
\times J_{m}(\frac{e V_{\alpha}}{\Omega_0}) {\cal G}_{j_{\alpha}, j_{\beta} }(k+q,\omega )
{\cal G}^*_{j_{\alpha}, j_{\beta} }(k,\omega )
  \Big]
 \Big\},
\label{jacrescte0}
\end{array}
\end{equation}
where we have used the first summation formula for products of Bessel
functions given in appendix E. Performing a shift
$\omega \rightarrow \omega - k \Omega_0$ in the second term of
(\ref{jacrescte0}) 
and making use of the identity (\ref{shoap0}), it is also possible to recast 
the above expression as:
\begin{eqnarray}
& & J_{\alpha}^{dc}   =    \sum_{q  m =-\infty}^{\infty} 
\mbox{Re} \Big\{ \int_{-\infty}^{\infty} \frac{d \omega}{2 \pi}  \Big[
 \Gamma^0_{\alpha}(\omega) \times
\nonumber \\
& &
 \sum_{\beta =1 }^{N_r}
\sum_{k=-\infty}^{\infty} \Big(
e^{i q \varphi_{\beta} } 
 J_{m+q}(\frac{eV_{\beta}}{\Omega_0}) J_{m}(\frac{eV_{\beta}}{\Omega_0})
\Gamma^0_{\beta}(\omega) \times    \nonumber \\
& & 
{\cal G}_{j_{\alpha}, j_{\beta} }(k-q,\omega +q \Omega_0)
{\cal G}^*_{j_{\alpha}, j_{\beta} }(k,\omega)  
f_{\beta}(\omega  -m  \Omega_0)  \Big)\nonumber \\
& &
- i e^{i q   \varphi_{\alpha} }  J_{m+q}(\frac{eV_{\alpha}}{\Omega_0}) J_{m}(\frac{eV_{\alpha}}{\Omega_0})
\Big({\cal G}_{j_{\alpha} j_{\alpha} }(q,\omega )  \nonumber \\
& & -
{\cal G}^*_{j_{\alpha} ,j_{\alpha} }(-q,\omega + q \Omega_0 ) \Big) 
 f_{\alpha}(\omega -m \Omega_0 ) \Big]
 \Big\}.
\label{jacrescte}
\end{eqnarray}
In summary, expressions (\ref{jacres0}) and (\ref{jacres}) define two equivalent
 ways to calculate the dc current through the lead $\alpha$ in the case of
 ac voltages at reservoirs with arbitrary densities of states, while 
(\ref{jacrescte0}) and (\ref{jacrescte}) are two equivalent representations of
 such current for reservoirs with smooth densities of states.

 In any of these cases, the evaluation of $J_{\alpha}^{dc}$ implies the  solution
 of the complete set (\ref{retfre}).

\section{Translation between the two formalisms}
We propose the following translation between the Floquet S-matrix and the
Green's functions:
\begin{eqnarray}
& & S_{F,\alpha \beta}(E_m,E_n)= 
\delta_{\alpha,\beta}\delta_{m-n,0}-  \\
& &
i \sqrt{\Gamma^0_{\alpha}( \omega +m\Omega_0) \Gamma^0_{\beta}(\omega +n\Omega_0)}
{\cal G}_{j_{\alpha}, j_{\beta}}(m-n,\omega + n\Omega_0),\nonumber
\label{trans}
\end{eqnarray}
with  $E_m= \omega +m \Omega_0$.

It is important to note that 
this translation exactly recovers all the representations for the dc current
derived in the framework of Floquet scattering matrix theory in the case of  
stationary reservoirs.
In fact, translating (\ref{ref}) and (\ref{curl}) according to (\ref{trans}),
leads to equations (\ref{smf_8}) and (\ref{smf_9}), respectively.
In addition,
in eq. (\ref{ref}) it is possible to identify the transfer matrix 
formulation of Refs.~\onlinecite{sols} and \onlinecite{flohan}:
\begin{equation}
T_{\alpha \beta}(E, E_k) =
|{\cal G}_{j_{\alpha}, j_{\beta} }(k,\omega)|^2 \Gamma^0_{\beta }(\omega)  
\Gamma^0_{\alpha}(\omega + k \Omega_0), 
\label{tran}
\end{equation} 
being $\alpha \neq \beta$. 

For stationary problems, we should consider $m=n=0$ 
and ${\cal G}_{j_{\alpha} j_{\beta}}(k ,\omega ) \rightarrow G^0_{j_{\alpha}, j_{\beta}} (\omega )$
in (\ref{trans}), in which
case, we recover an expression like the one proposed in Ref.~\onlinecite{file}:
\begin{eqnarray}
& & S^0_{\alpha \beta}(E)= 
\delta_{\alpha,\beta} - \nonumber \\
& &
i \sqrt{\Gamma^0_{\alpha}( \omega) \Gamma^0_{\beta}(\omega )}
G^0_{j_{\alpha}, j_{\beta}}(\omega ),
\label{trans0}
\end{eqnarray}

In the case of reservoirs with ac-voltages  described by  
wide-band models with smooth densities of states, such that
$\Gamma_{\alpha}(\omega \pm k \Omega_0) \sim  \Gamma_{\alpha}(\omega )$, 
the translation (\ref{trans}) also transforms Eq.(\ref{jacrescte}) into
Eq.(\ref{smf_13}). 

\section{Proof of the unitary property of $\hat{S}_F$.}
In the S-matrix formalism, current conservation implies that the scattering matrix is unitary,
see Eqs.(\ref{smf_4}).
%
In what follows we show that, for static reservoirs,
this  property can be proved by using 
identities satisfied by the Green's functions as well as the translation formula
(\ref{trans}). In fact,
let us use Eq.(\ref{trans}) inside the summations of the left hand side of Eq. (\ref{smf_4B}):

\begin{eqnarray}
 & & \sum_{\beta=1}^{N_r} \sum_{n=-\infty}^{\infty}
S^*_{F,\alpha \beta}(E,E_n)S_{F,\gamma \beta}(E_m,E_n) =
\nonumber \\
& & \delta_{\alpha \gamma} \delta_{m,0} - 
\sqrt{ \Gamma^0_{\gamma}( \omega +m \Omega_0)
  \Gamma^0_{\alpha}(\omega) }\Big[ i{\cal G}_{j_{\gamma}, j_{\alpha}}(m,
\omega) -
\nonumber \\
& &
 i{\cal G}^*_{j_{\alpha}, j_{\gamma} }(-m,\omega+m \Omega_0) 
- \sum_{\beta=1}^{N_r} \sum_{n=-\infty}^{\infty} {\cal G}_{j_{\gamma},
  j_{\beta}}(m-n,\omega+n\Omega_0)
\nonumber \\
& &
\times
\Gamma^0_{\beta}(\omega+n\Omega_0) {\cal G}^*_{j_{\alpha}, j_{\beta} }(-n,\omega+n \Omega_0) \Big].
\end{eqnarray}
From the identity (\ref{shoap}), it can be shown that the second term
of the right hand side of the above equation vanishes identically, thus
recovering equation (\ref{smf_4B}).

Similarly, using Eq.(\ref{trans}) inside the summations of the left hand side of
Eq. (\ref{smf_4A}):
\begin{eqnarray}
& &  \sum_{\alpha=1}^{N_r} \sum_{n=-\infty}^{\infty}
S^*_{F,\alpha \beta}(E_n,E) S_{F, \alpha \gamma }(E_n,E_m) =
\nonumber \\
& &
\delta_{\beta \gamma} \delta_{m,0} +  
\sqrt{ \Gamma^0_{\gamma}(\omega+m \Omega_0)
  \Gamma^0_{\beta}(\omega) } \Big[ i{\cal G}_{j_{\gamma}, j_{\beta}}^*(m,\omega) - 
\nonumber \\
& &
 i{\cal G}_{ j_{\beta}, j_{\gamma} }(-m,\omega+m \Omega_0) 
+ \sum_{\alpha=1}^{N_r} \sum_{n=-\infty}^{\infty} {\cal G}_{j_{\alpha},
  j_{\beta}}^*(n,\omega)
\nonumber \\
& & \times
\Gamma^0_{\alpha}(\omega+n\Omega_0) 
{\cal G}_{j_{\alpha}, j_{\gamma}  }(-m+n,\omega+m \Omega_0) \Big], 
\end{eqnarray}
and from the identity (\ref{shoap1}), equation (\ref{smf_4A}) is
recovered.

The same procedure can be followed to prove the unitarity of the S-matrix in the case of reservoirs with oscillating
voltages at the reservoirs, provided the density of states of the reservoirs is smooth.

\section{Adiabatic approximation within Keldysh formalism.}
\subsection{Definition}
The adiabatic point of view is inspired in a parametrical
representation of the time dependent terms of the Hamiltonian. This means a description where
the observation time $t$ is assumed to be frozen in the
equations governing the dynamics of the system. In particular,
 instead of the Dyson equation (\ref{retfre}), in a frozen description we must
 consider the following equation
\begin{eqnarray}
& & \hat{G}^{f } (t,\omega) = \hat{G}^0(\omega) + \nonumber \\
& & {\sum_{k=-\infty}^{\infty}}^{\prime} e^{-i k \Omega_0 t } 
\hat{G}^{f }(t,\omega) \hat{V}(k) \hat{G}^0(\omega)
+ \nonumber \\
& & {\sum_{k=-\infty}^{\infty}}^{\prime}  e^{-i k \Omega_0 t} 
\hat{G}^{f}(t,\omega) \hat{\Sigma}(k,\omega) \hat{G}^0(\omega)
\label{frodyre}
\end{eqnarray}
which is a stationary Dyson equation corresponding to the strength of the
parameters 
$\hat{V}(t)$ and $V_{\alpha}(t)$
at the observation time $t$. As the potentials are periodic
in $t$, the frozen Green's function can be expanded in a Fourier series:
\begin{equation}
 \hat{G}^{f}(t,\omega) = \sum_{k=-\infty}^{\infty} e^{-ik \Omega_0 t} 
\hat{\cal G}^f (k,\omega),
\end{equation}
and through the translation (\ref{trans0}) it is possible to define the Fourier coefficients 
$\hat S_0(k,E)\equiv \hat S_{0,k}(E)$ for the elements of the frozen S-matrix as
 \begin{equation}
S_{0,\alpha \beta}(k,E)= 
\delta_{\alpha,\beta}\delta_{k,0}-
i \sqrt{\Gamma^0_{\alpha}(\omega) \Gamma^0_{\beta}(\omega) }
{\cal G}^f_{j_{\alpha}, j_{\beta}}(k,\omega).
\label{statio}
\end{equation}

In the  Floquet S-matrix formalism, the adiabatic approximation is given by Eqs.(\ref{smf_15}).
In analogy to Eq.(\ref{smf_15A}) we propose the
following ansatz for the $\propto \Omega_0$ approximation to the Green's
function:
\begin{equation}
\hat{\cal G}^R(k, \omega) \sim \hat{\cal G}^f(k, \omega)+
\frac{k \Omega_0}{2} \frac{ \partial \hat{\cal G}^f(k, \omega)}
{\partial \omega} + \Omega_0 \hat{a}(k,\omega),
\label{greenad}
\end{equation}
or, equivalently,
\begin{equation}
\hat{G}(t, \omega) \sim \hat{G}^f(t, \omega)+i
\frac{1}{2} \frac{ \partial^2  \hat{G}^f(t, \omega)}
{\partial t \partial \omega } + \Omega_0 \hat{a}(t,\omega),
\label{greenadt}
\end{equation}
where $\hat{\cal G}^f(k, \omega)$ (or $\hat{G}^f(t, \omega)$) is the {\em frozen} Green's function,
which obeys the equilibrium Dyson equation (\ref{frodyre}).
The substitution of the ansatz Eq.(\ref{greenad}) with the translation formula
(\ref{trans}) into Eq.(\ref{smf_15A}) leads to the
following relation between $A_{\alpha \beta}$ and $a_{j_{\alpha}, j_{\beta}}$:
\begin{eqnarray}
& &A_{\alpha \beta}(k,E) = -i \sqrt{\Gamma^0_{\alpha}(\omega) \Gamma^0_{\beta}(\omega) }
a_{j_{\alpha}, j_{\beta}}(k,\omega) - \nonumber \\
& & 
i \frac{k \Omega_0}{4} \sqrt{\Gamma^0_{\alpha}(\omega) \Gamma^0_{\beta}(\omega) }
{\cal G}^f_{j_{\alpha}, j_{\beta}}(k, \omega) \nonumber \\
& &
\Big[ \frac{1}{\Gamma^0_{\alpha}(\omega)} 
\frac{\partial \Gamma^0_{\alpha}(\omega)}{\partial \omega}-
 \frac{1}{\Gamma^0_{\beta}(\omega)} \frac{\partial \Gamma^0_{\beta}(\omega)}{\partial
  \omega} \Big],
\nonumber \\
& & A_{\alpha \beta}(t,E) = -i 
\sqrt{\Gamma^0_{\alpha}(\omega) \Gamma^0_{\beta}(\omega) }
a_{j_{\alpha} , j_{\beta}}(t,\omega) +\nonumber \\
& & 
 \frac{1 }{4} \sqrt{\Gamma^0_{\alpha}(\omega) \Gamma^0_{\beta}(\omega) }
\frac{ \partial G^f_{j_{\alpha} , j_{\beta}}(t, \omega) }
{\partial t} \nonumber \\
& &
\Big[ \frac{1}{\Gamma^0_{\alpha}(\omega)} \frac{\partial \Gamma^0_{\alpha}(\omega)}{\partial
  \omega}-
 \frac{1}{\Gamma^0_{\beta}(\omega)} \frac{\partial \Gamma^0_{\beta}(\omega)}{\partial
  \omega} \Big].
\label{aa}
\end{eqnarray}

In  Ref.~\onlinecite{micha3}, some important properties of the matrix $\hat{A}$ have been
proved on the basis of the unitary property of $\hat{S}_F$ and the fact that
for a Hamiltonian of spinless fermions  that depends on a magnetic flux $\Phi$ and
on time-dependent potentials of the form 
\begin{equation}
V_{l,l'}(t)= \delta_{l, l'} \sum_j \delta_{l,j}
[V_j^0 + V^1_j \cos(\omega t + \delta_j)],
\label{pot}
\end{equation}
the stationary matrix transforms under  $t \rightarrow -t$ as:
\begin{equation}
S_{0, \alpha \beta}(k,E,\Phi,\delta)= S_{0, \alpha \beta}(-k,E,\Phi,-\delta),
\end{equation}
as well as
\begin{equation}
S_{0, \alpha \beta}(k,E,-\Phi,\delta)= S_{0, \beta \alpha }(k,E,\Phi,\delta).
\end{equation}

Analogously, the Hamiltonian is invariant under the simultaneous
change of $t \rightarrow -t$ and $\delta_j \rightarrow -\delta_j$. Therefore,
\begin{equation}
{\cal G}^{f}_{l,l'}(k,\Phi,\delta_j,\omega)={\cal G}^{f }_{l,l'}(-k,\Phi,-\delta_j,\omega).
\end{equation}
In addition, in the presence of a magnetic flux $\Phi$, the static terms of the
 Hamiltonian of the system
satisfy $\varepsilon_{l ,l'}(\Phi)= \varepsilon_{l',l}(-\Phi)=
[\varepsilon_{l', l}(\Phi)]^*$, which implies:
\begin{equation}
G^{f}_{l,l'}(t,\Phi,\omega)=G^{f}_{l',l}(t,-\Phi,\omega).
\end{equation}
In other words, the frozen Green's function ${\cal G}^f_{j_{\alpha},j_{\beta}
 }(k, \omega)$ has
 the same
symmetry properties of $S_{0,\alpha\beta }(k, \omega)$. Therefore, from Eq. (\ref{aa}) we see that
$a_{j_{\alpha} j_{\beta}} $ has the same symmetry properties as $A_{\alpha \beta }$.

In order to calculate $\hat{a}(k, \omega)$ explicitly, we have 
to consider the Dyson equation for the Fourier coefficients of
the retarded Green's function, which can be obtained by expanding 
(\ref{retfre}) in Fourier series:
\begin{eqnarray}
& & \hat{\cal G}(k,\omega) = \hat{G}^0(\omega) \delta_{k,0} + \nonumber \\
& &  {\sum_{k'=-\infty}^{\infty}}^{\prime}
 \hat{\cal G}(k+k',\omega+k' \Omega_0) \hat{V}(k')  \hat{G}^0(\omega) + \nonumber \\
& &  {\sum_{k'=-\infty}^{\infty}}^{\prime}
 \hat{\cal G}(k+k',\omega+k' \Omega_0) \hat{\Sigma} (k',\omega)  \hat{G}^0(\omega),
\label{retfou}
\end{eqnarray}
then substitute (\ref{greenad}) 
 and keep terms up to the first order in $\Omega_0$.
The solution of the ensuing linear set allows for the evaluation of 
$\hat{a}(k,\omega)$ as a function of the stationary Green functions
$\hat{G}^0(\omega)$, the frozen Green functions $\hat{\cal G}^f(k,\omega)$
and the derivatives $\partial \hat{\cal G}^f (k,\omega)/\partial \omega$
and  $\partial \hat{G}^0(\omega)/\partial \omega$.
Alternatively, in order to get $\hat{a}(t,\omega)$, we have to substitute the
ansatz (\ref{greenadt}) in the Dyson equation (\ref{retfre}),
keep terms up to $\propto \Omega_0$ 
and solve the resulting linear set, which  gives $\hat{a}(t, \omega)$
as a function of $\hat{G}^{f  }(t,\omega)$, 
$\partial^2 \hat{G}^{f  }(t,\omega)/\partial t \partial \omega$,
$\hat{G}^0 (\omega)$,
and $\partial \hat{G}^0 (\omega)/\partial \omega$.

\subsection{Calculation of the dc current}
\subsubsection{Time-periodic local potentials with stationary reservoirs}
In order to calculate the adiabatic approximation to the
 dc-current through the leads in the case of stationary
reservoirs it is convenient to start from Eq. 
(\ref{curl}). Expanding that expression in powers of $\Omega_0$ and keeping
up to the linear term, the dc-current reads:
\begin{eqnarray}
& & J^{dc}_{\alpha} 
\sim  \frac{1}{\tau_0} \int_{0}^{\tau_0} dt \int_{-\infty}^{\infty} \frac{d \omega}{2 \pi}  \Big[
\frac{\partial f_{\alpha}(\omega) }{\partial \omega } J^{(pump)}_{\alpha}(t,\omega) +
\nonumber \\
& &   \Big( f_{\beta}(\omega) - f_{\alpha}(\omega) \Big)
\Big( J^{(bias)}_{\alpha}(t,\omega) + J^{(int)}_{\alpha}(t,\omega)\Big)  \Big],
\end{eqnarray}
being 
\begin{subequations}
\begin{eqnarray}
& & J^{(pump)}_{\alpha}(t,\omega) =  \Gamma_{\alpha}^0(\omega)
 \Gamma_{\beta}^0(\omega) \times\nonumber \\
& &
 \sum_{\beta =1 }^{N_r} \frac{1}{2} \Big[ i G^f_{j_{\alpha}, j_{\beta} }(t, \omega)
 \frac{\partial G^f_{j_{\alpha}, j_{\beta}}(t, \omega)^*}{\partial t} 
+ c.c. \Big] 
, \label{pump}
\end{eqnarray} 
\begin{equation}
 J^{(bias)}_{\alpha}(t,\omega)= \sum_{\beta =1 }^{N_r} |G^f_{j_{\alpha},
  j_{\beta} }(t, \omega)|^2  \Gamma^0_{\alpha}(\omega) \Gamma^0_{\beta}(\omega),
\label{bias}
\end{equation}
\begin{eqnarray}
& & J^{(int)}_{\alpha}(t,\omega) =  \sum_{\beta =1 }^{N_r} \frac{1}{2}
\frac{\partial \Gamma^0_{\alpha}(\omega)}{\partial\omega}
\Gamma^0_{\beta}(\omega) \times
\nonumber \\
& &
\Big[ -
 i G^f_{j_{\alpha}, j_{\beta} }(t, \omega)
 \frac{\partial G^f_{j_{\alpha}, j_{\beta}}(t, \omega)^*}{\partial t} 
+ c.c. \Big] \nonumber \\
& &
  + \frac{\Gamma^0_{\alpha}(\omega) \Gamma^0_{\beta}(\omega)}{2} \Big[ - i \frac{\partial G^f_{j_{\alpha}, j_{\beta} }(t, \omega)}{\partial \omega}
 \frac{\partial G^f_{j_{\alpha}, j_{\beta}}(t, \omega)^*}{\partial t} +
\nonumber \\
& & \Omega_0 \Big( G^f_{j_{\alpha}, j_{\beta}}(t, \omega) a_{j_{\alpha}, j_{\beta}}(t,\omega) 
+ c.c. \Big) \Big],
\label{int}
\end{eqnarray}
\end{subequations}
For the case of reservoirs with identical chemical potentials and
 temperatures, such that $f_{\alpha}(\omega)=f_0(\omega), \forall \alpha$, only the pumping
 term (\ref{pump})
contributes. This term is $\propto \Omega_0$ and can be shown to be equivalent
 to Eq.(\ref{smf_18B}) through the
 translation formula (\ref{trans}). The second term is the usual stationary contribution
(\ref{curest}). The ``interference'' term ($\propto \Omega_0 (\mu_{\beta} - \mu_{\alpha})$)
is a small contribution in the limit of small static bias $\mu_{\beta} -
 \mu_{\alpha}$.
However, it may give rise to interesting behavior in the presence of magnetic  field. 
In particular, in a two terminal set-up it is an odd function of a magnetic field~~\cite{micha3},  in striking contrast with a stationary conductance which is an even function of a magnetic field.

\subsubsection{Time-dependent voltages at the reservoirs.}
In order to derive the $\propto \Omega_0$ contribution to the dc current, we
substitute the adiabatic approximation to the Green's function (\ref{greenad})
in (\ref{jacres0}) and we expand the remaining terms up to 
${\cal O}(\Omega_0^2)$. The latter step results in the following expansion:
\begin{eqnarray}
& & \Gamma_{\alpha}(p,\omega+k \Omega_0) \Gamma^<_{\beta}(q,\omega)
-
\Gamma^<_{\alpha}(p,\omega+k \Omega_0) \Gamma_{\beta}(q,\omega)  
\nonumber \\
& &
\sim \sum_{m,n-\infty}^{\infty} J_{n+p}(\frac{eV_{\alpha}}{\Omega_0})  J_{n}(\frac{eV_{\alpha}}{\Omega_0}) 
J_{m+q}(\frac{eV_{\beta}}{\Omega_0})  J_{m}(\frac{eV_{\beta}}{\Omega_0}) 
\nonumber \\
& & \times e^{i (p \varphi_{\alpha} + q \varphi_{\beta}) }
   \Big\{ \Gamma_{\alpha}(\omega) \Gamma_{\beta}(\omega)
\Big[ f_{\beta}(\omega)-f_{\alpha}(\omega) \Big]+
\nonumber \\
& &
\Omega_0 \Big[ (k-n) g_1(\omega) -m g_2(\omega)\Big]
+ \Omega_0^2 \Big[g_3(\omega)(k-n)^2 
\nonumber \\
& &
+ m^2 g_4(\omega) + m(n-k) g_5(\omega) \Big] \Big\},
\label{expand}
\end{eqnarray}
being
\begin{eqnarray}
g_1(\omega)& = & \Gamma_{\beta}(\omega) \Big\{ 
\frac{\partial  \Gamma_{\alpha}(\omega)}{\partial \omega} 
\Big[ f_{\beta}(\omega)-f_{\alpha}(\omega) \Big] \nonumber \\
& &
- \Gamma_{\alpha}(\omega)
\frac{\partial f_{\alpha}(\omega) }{\partial \omega} \Big\}, \nonumber \\
g_2(\omega)& = & \Gamma_{\alpha}(\omega) \Big\{ 
\frac{\partial  \Gamma_{\beta}(\omega)}{\partial \omega} 
\Big[ f_{\beta}(\omega)-f_{\alpha}(\omega) \Big] 
\nonumber \\
& &
+ \Gamma_{\beta}(\omega)
\frac{\partial f_{\beta}(\omega) }{\partial \omega} \Big\}, \nonumber \\
g_3(\omega)& = & \frac{\Gamma_{\beta}(\omega)}{2} \Big\{ 
\frac{\partial^2  \Gamma_{\alpha}(\omega)}{\partial \omega^2} 
\Big[ f_{\beta}(\omega)-f_{\alpha}(\omega) \Big] 
\nonumber \\
& &
- \Gamma_{\alpha}(\omega)
\frac{\partial^2 f_{\alpha}(\omega) }{\partial \omega^2} \Big\}, \nonumber \\
g_4(\omega)& = & \frac{\Gamma_{\alpha}(\omega)}{2} \Big\{ 
\frac{\partial^2  \Gamma_{\beta}(\omega)}{\partial \omega^2} 
\Big[ f_{\beta}(\omega)-f_{\alpha}(\omega) \Big] 
\nonumber \\
& &
+ \Gamma_{\beta}(\omega)
\frac{\partial^2 f_{\beta}(\omega) }{\partial \omega^2} \Big\}, \nonumber \\
g_5(\omega)& = & \frac{\partial  \Gamma_{\alpha}(\omega)}{\partial \omega} 
\frac{\partial  \Gamma_{\beta}(\omega)}{\partial \omega}
\Big[ f_{\beta}(\omega)-f_{\alpha}(\omega) \Big].
\label{defig}
\end{eqnarray}
When substituting in (\ref{jacres0})
we  use the properties of the Bessel function enunciated in appendix E and
 keep only terms proportional to $\Omega_0$ and
$V_{\alpha}$. The resulting expression shows a rather compact form in the 
case of reservoirs with smooth densities of states and  the same chemical
potentials and temperature,
 such that $f_{\alpha}(\omega)= f_0(\omega), \forall \alpha$, in which case $g_5(\omega)=0$,
while $g_1(\omega)= - g_2 (\omega) $ and $g_3(\omega)= - g_4 (\omega) $.
Terms $\propto \partial^2 f_0(\omega)/\partial \omega^2$ can be reduced to
terms $\propto \partial f_0(\omega)/\partial \omega$ by integrating by parts.
The final result can be written by collecting the different terms
in three  kinds of  contributions
as in Eq.(\ref{smf_18}): 
\begin{eqnarray}
& & J^{dc}_{\alpha} \sim  \frac{1}{\tau_0} \int_{0}^{\tau_0} dt
\int_{-\infty}^{\infty}
\frac{d \omega}{2 \pi} \frac{\partial f_0(\omega) }{\partial \omega }
\nonumber \\
& & \times \Big[  J^{(pump)}_{\alpha}(t,\omega) +  J^{(rect)}_{\alpha}(t,\omega) +
 J^{(int)}_{\alpha}(t,\omega) \Big],
\end{eqnarray}
with $J^{(pump)}_{\alpha}(t,\omega)$ given by (\ref{pump}) and
\begin{subequations}
\begin{eqnarray}
& & J^{(rect)}_{\alpha}(t,\omega) = \sum_{\beta =1 }^{N_r}  G^f_{j_{\alpha}, j_{\beta}}(t, \omega)
G^f_{j_{\alpha}, j_{\beta}}(t, \omega)^*  
\nonumber \\
& &
\times
 \Gamma_{\alpha}(\omega) \Gamma_{\beta}(\omega)
\Big[ V_{\alpha}(t)- V_{\beta}(t) \Big], 
\label{rect}
\end{eqnarray}
\begin{eqnarray}
& & J^{(int)}_{\alpha}(t,\omega) =
 \sum_{\beta =1 }^{N_r}  \Gamma_{\alpha}(\omega) \Gamma_{\beta}(\omega)
\times \nonumber \\
& &
 V_{\beta}(t) \Big\{ \frac{1}{2} \Big[ i \frac{\partial G^f_{j_{\alpha}, j_{\beta} }(t,\omega) }{\partial \omega} 
\frac{\partial G^f_{j_{\alpha}, j_{\beta} }(t,\omega)^* }{\partial t} + c.c \Big] \nonumber \\
& & + 2 \Omega_0   \Big[  G^f_{j_{\alpha}, j_{\beta} }(t,\omega)
a_{j_{\alpha}, j_{\beta} }(t,\omega)^ * + c.c. \Big] \Big\}. \label{inter}
\end{eqnarray}
\end{subequations}
Through the translation formula (\ref{trans}), we can identify the
three terms (\ref{pump}), (\ref{rect}) and (\ref{inter}) with (\ref{smf_18B}),
(\ref{smf_18C}) and (\ref{smf_18D}), obtained with the S-matrix formalism.
In the derivation of the expression for the interference term (\ref{inter})
we have made use of (\ref{shoap0}) expressed in the adiabatic approximation
(\ref{greenad})
and an equivalent equation satisfied by
the frozen Green function:
\begin{eqnarray}
& & {\cal G}^f_{j_{\alpha}, j_{\alpha}} (k, \omega)-
{\cal G}^f_{j_{\alpha}, j_{\alpha}} (k, \omega)^* = \nonumber \\
& &  -i \sum_{k'\beta} 
{\cal  G}^f_{j_{\alpha}, j_{\beta}} (k+ k', \omega) \Gamma^0_{\beta}(\omega) 
{\cal  G}^f_{j_{\alpha}, j_{\beta}} (k', \omega)^*,
\end{eqnarray}
which leads to the following relation:
\begin{eqnarray}
& & \frac{k}{2} \sum_{\beta =1 }^{N_r} \sum_{ k'=-\infty}^{\infty} \Gamma^0_{\beta}(\omega)
{\cal G}^f_{j_{\alpha}, j_{\beta}} (k + k', \omega) 
\frac{ \partial {\cal G}^f_{j_{\alpha}, j_{\beta}} (k', \omega)^*}{ \partial
  \omega} \nonumber \\
& & = \sum_{\beta  =1 }^{N_r} \sum_{ k'=-\infty}^{\infty} \Gamma^0_{\beta}(\omega) \Big\{ 
{\cal G}^f_{j_{\alpha}, j_{\beta}} (k+k', \omega) 
a_{j_{\alpha}, j_{\beta}}(k', \omega)^* \nonumber \\
& &
+ 
a_{j_{\alpha}, j_{\beta}}(k+k', \omega) {\cal G}^f_{j_{\alpha}, j_{\beta}}(k',
\omega)^*
\nonumber \\
& & - \frac{k'}{2} \Big[ 
 {\cal G}^f_{j_{\alpha}, j_{\beta}} (k+k', \omega)
\frac{ \partial {\cal G}^f_{j_{\alpha}, j_{\beta}} (k', \omega)^*}{ \partial
  \omega} + 
\nonumber \\
& &
\frac{ \partial{\cal G}^f_{j_{\alpha}, j_{\beta}} (k+k', \omega) }{ \partial \omega}
 {\cal G}^f_{j_{\alpha}, j_{\beta}} (k', \omega)^* \Big] \Big\}.
\end{eqnarray}

\section{Summary and conclusions}
Starting from the formulation of Keldysh non-equilibrium 
approach to systems in the presence of time-periodic fields of
Ref. ~\onlinecite{lilip}, we have shown
 several identities satisfied the Green's function. 
This has allowed us for the derivation of several useful equations to
calculate the current (in particular, the dc-component of the current) flowing between
the reservoirs and the mesoscopic system. We have considered two different
situations: (i) Driving induced by voltages applied at the central
structure. (ii) Driving induced by voltages applied at the reservoirs. In both
cases we have considered reservoirs with a general  density of states.  
We have also proposed an expression
that enables the translation between Green's function formalism and the Floquet
Scattering
matrix formalism of Refs. ~\onlinecite{micha1, micha2, micha3}. In the case
(i), we have shown 
that this formula is able to translate exactly the expressions for the dc current
flowing through the leads obtained in the two formalisms. Furthermore, we have
shown that it is enough to derive the unitary property of the scattering
matrix by recourse to properties of the Green's functions. In the situation (ii)
we were also able to translate  expressions for the current and to demonstrate
the unitary property of $S_F$ if we assume within the Green's function
formalism models of reservoirs with a smooth density of states. 

We have also formulated the so called adiabatic approximation to the
dc-current in the framework of the non-equilibrium Green's function formalism.
We have used it to derive the different contributions to the dc-current linear in the 
driving frequency in the two situations (i) and (ii) described above. 
Making use of the translation formula of section IV, it is possible to compare
these expressions with the ones previously derived in the framework of Scattering
matrix theory \cite{micha1,micha2,micha3}. The equivalence is complete in the
cases of stationary reservoirs as well as in the case of oscillating
reservoirs with a smooth density of states.

\section{Acknowledgments}
LA thanks Victor Gopar for the careful reading of this manuscript and
constructive comments.
Support from
PICT 03-11609 from Argentina, BFM2003-08532-C02-01
from MCEyC of Spain, grant ``Grupo de investigaci\'on de excelencia DGA''
and from the MCEyC of Spain through ``Ramon y Cajal'' program
 are
acknowledged. LA is staff member of CONICET, Argentina.

\appendix

\section{ }
In the stationary system with $V_{l,l'}(t)=V_{\alpha}(t)=0$, the 
Dyson equation for the retarded Green function is simply Eq. (\ref{dy0}).
From there, we can write
\begin{eqnarray}
& & \hat{G}^0(\omega) - \hat{G}^0(\omega)^{\dagger}  = 
\nonumber \\
 & = &  \Big\{ \hat{1} \omega - \hat{\varepsilon} - 
\hat{\Sigma}^0(\omega) \Big\}^{-1} - 
 \Big\{ \hat{1} \omega - \hat{\varepsilon}^{\dagger}- 
\hat{\Sigma}^0(\omega)^{\dagger} \Big\}^{-1} \nonumber \\
& = &  
\Big\{ \hat{1} \omega - \hat{\varepsilon}- \hat{\Sigma}^0(\omega) \Big\}^{-1} 
\nonumber \\
& & \times 
\Big\{ \hat{1} \omega - \hat{\varepsilon}^{\dagger} - \hat{\Sigma}^0(\omega)^{\dagger} \Big\}
\Big\{ \hat{1} \omega - \hat{\varepsilon}^{\dagger} - \hat{\Sigma}^0(\omega)^{\dagger} \Big\}^{-1}
\nonumber \\
& - &
\Big\{  \hat{1} \omega - \hat{\varepsilon}- 
\hat{\Sigma}^0(\omega) \Big\}^{-1} \Big\{ \hat{1} \omega - \hat{\varepsilon}
 - \hat{\Sigma}^0(\omega) \Big\}
\nonumber \\
& & \times
\Big\{ \hat{1} \omega - \hat{\varepsilon}^{\dagger} - \hat{\Sigma}^0(\omega)^{\dagger} \Big\}^{-1} 
\nonumber \\
& = &  \hat{G}^0(\omega) \Big\{ \hat{\Sigma}^0(\omega)-\hat{\Sigma}^0(\omega)^{\dagger} 
\Big\}\hat{G}^0(\omega)^{\dagger},
\end{eqnarray}
where we have used the fact that  $\hat{\varepsilon}=\hat{\varepsilon}^{\dagger}$.

\section{ }
Let us start from the definition of the dc current flowing through the lead $\alpha$, Eq.(\ref{jl}).
The condition of the continuity of the current implies:
\begin{eqnarray}
& & \sum_{\alpha=1 }^{N_r} J_{\alpha}^{dc}=0 \nonumber \\
& &=  2 \sum_{\alpha =1 }^{N_r} \int_{-\infty}^{\infty} \frac{d \omega}{2 \pi}
 \mbox{Im}\Big[{\cal G}_{j_{\alpha}, j_{\alpha} } (0, \omega)\Big] f_{\alpha}(\omega) +
 \sum_{\alpha=1 }^{N_r} \sum_{k=-\infty}^{\infty} 
 \nonumber \\
& & \int_{-\infty}^{\infty}
 \frac{d \omega}{2 \pi} 
 f_{\beta}(\omega) |{\cal G}_{j_{\alpha}, j_{\beta} }(k, \omega) |^2
\Gamma^0_{\beta}(\omega) \Gamma^0_{\alpha}(\omega + k \Omega_0) .
\end{eqnarray}
Since the above equation must hold for arbitrary chemical potentials and
temperatures of the reservoirs,
the 
terms multiplying the Fermi functions  $f_{\alpha}(\omega)$ 
  have to vanish for each $\alpha$, which means:
\begin{equation} 
-2 \mbox{Im}\Big[{\cal G}_{j_{\alpha}, j_{\alpha}}(0, \omega)\Big] =
\sum_{\beta =1 }^{N_r} \sum_{k=-\infty}^{\infty}
|{\cal G}_{j_{\beta}, j_{\alpha}} (k, \omega) |^2 
\Gamma^0_{\beta}(\omega + k \Omega_0).
\label{idcont}
\end{equation}

\section{ }
We present the proof of an important identity for the retarded
Green's function. This identity has been previously proved for the
case of static reservoirs in Ref.~\onlinecite{flohan}
within the framework of a different formalism.  

We first perform a Fourier transform
in $t-t'$ in the first equation of the set (\ref{difdyr}) and take the
adjoint in the spacial indices of that equation. The result is:
\begin{eqnarray}
& & \Big\{ i \frac{\overrightarrow{\partial}}{\partial t} + (\omega' + i
0^+)-\hat{H}^{sys}(t) \Big\}
\hat{G}^R(t,\omega') \nonumber \\
& &
- \int_{-\infty}^t dt_1 e^{i \omega' (t-t_1)}
\hat{\Sigma}^R(t,t_1) \hat{G}(t_1,\omega')
=  \hat{1},  \label{a1} \\
& & 
\hat{G}^R(t,\omega)^{\dagger} \Big\{ -i 
\frac{\overleftarrow{\partial}}{\partial t} + (\omega - i 0^+)-
\hat{H}^{sys}(t)  \Big\} \nonumber \\
& &
 - \int_{-\infty}^t  dt_1 e^{-i \omega (t-t_1)}
 \hat{G}^R(t_1,\omega)^{\dagger} \hat{\Sigma}^R(t,t_1)^{\dagger}
=  \hat{1}. \label{a2}
\end{eqnarray}
 Note that, unlike the stationary case, these
equations are not simplified to a linear set of equations and the inverse
of the Green's function must be represented in terms of integro-differential
operators.
By multiplying (\ref{a1}) from the left  by $\hat{G}^{R}(t,\omega)^{\dagger}$
and (\ref{a2}) from the right by $\hat{G}^R(t,\omega')$ and then subtracting
the two resulting equations we get
\begin{eqnarray}
& & \hat{G}^R(t,\omega)^{\dagger} - \hat{G}^R(t,\omega') =
i \frac{\partial}{\partial t} \Big[\hat{G}^R(t,\omega)^{\dagger}
\hat{G}^R(t,\omega') \Big]+
\nonumber \\
& & 
(\omega'-\omega) \hat{G}^R(t,\omega)^{\dagger} \hat{G}^R(t,\omega') -
\nonumber \\
& &\int_{-\infty}^t  dt_1 e^{i \omega' (t-t_1)} \hat{G}^R(t,\omega)^{\dagger}
\hat{\Sigma}^R(t,t_1)
 \hat{G}^R (t_1,\omega')+
\nonumber \\
& & 
 \int_{-\infty}^t  dt_1 e^{- i \omega (t-t_1)} \hat{G}^R(t_1,\omega)^{\dagger} 
\hat{\Sigma}^R(t,t_1)^{\dagger} 
\hat{G}^R(t,\omega').
\end{eqnarray}
If we now perform the expansions in Fourier series for $\hat{G}^R(t,\omega)$
and  $\hat{G}^R(t,\omega')^{\dagger}$, we obtain:
\begin{eqnarray}
& & \hat{\cal G}(-k, \omega)^{\dagger} - \hat{\cal G}(k, \omega')  = 
\nonumber \\
& & 
(\omega' - \omega+ k \Omega_0) 
\sum_{k'=-\infty}^{\infty} \hat{\cal G}(k', \omega)^{\dagger}\hat{\cal G}(k+k', \omega')-
\nonumber \\
& & \sum_{k', k''=-\infty}^{\infty} \hat{\cal G}^{\dagger}(k', \omega) 
 \Big\{ \hat{\Sigma}(k'', \omega' + (k+k'-k'') \Omega_0) - \nonumber \\
& & 
\hat{\Sigma}(k'',\omega + k' \Omega_0)^{\dagger} \Big\}  \hat{\cal G}(k+k'-k'', \omega').
\end{eqnarray}
For $\omega'=\omega-k \Omega_0$, the above equation reduces to
\begin{eqnarray}
& & \hat{\cal G}(-k, \omega)^{\dagger} - \hat{\cal G}(k, \omega-k \Omega_0)  = 
\nonumber \\
& &
\sum_{k', k''} \hat{\cal G}(k', \omega)^{\dagger} 
\Big\{
\hat{\Sigma}^{\dagger}(k'',\omega + (k'-k'') \Omega_0)
- 
\nonumber \\
& &
\hat{\Sigma}(k'',\omega + k' \Omega_0) 
\Big\}  \hat{\cal G}(k+k'-k'', \omega-k \Omega_0).
\label{shoapi1}
\end{eqnarray}
In the case of stationary reservoirs, the above equation further reduces to
\begin{eqnarray}
& & \hat{\cal G}(-k, \omega)^{\dagger} - \hat{\cal G}(k, \omega-k \Omega_0)  = 
\nonumber \\
& &
\sum_{k'=-\infty}^{\infty} \hat{\cal G}(k', \omega)^{\dagger} 
\Big[
\hat{\Sigma}^0(\omega + k' \Omega_0)^{\dagger}
- 
\nonumber \\
& &
\hat{\Sigma}^0(\omega + k' \Omega_0) \Big]  \hat{\cal G}(k+k', \omega-k \Omega_0),
\label{shoap1}
\end{eqnarray}
which, for $k=0$, reads: 
\begin{eqnarray}
& & \hat{\cal G}(0, \omega) - \hat{\cal G}(0, \omega)^{\dagger}  =  
\sum_{k=-\infty}^{\infty} \hat{\cal G}(k, \omega)^{\dagger} 
\Big[ \hat{\Sigma}^0(\omega+k\Omega_0) - 
\nonumber \\
& & \hat{\Sigma}^0(\omega+k\Omega_0)^{\dagger} \Big ] \hat{\cal G}(k, \omega).
\label{shoap10}
\end{eqnarray}
Note that the identity (\ref{idcont}) derived in the previous appendix
 is a particular case of this equation.

The above identities are also valid in the case of oscillating reservoirs with
smooth densities of states, as can be verified by using the fact that
$\Gamma_{\alpha}^0(\omega + k \Omega_0) \sim \Gamma_{\alpha}^0(\omega )$ 
in (\ref{sigma}) as well as the first of the summation formulas of products of
Bessel functions of appendix E.

\section{ }
In this appendix we prove another important identity satisfied by the
Green's function.
We start from the definitions (\ref{gret}) and (\ref{gles})
of the different Green's functions in Keldysh
formalism and 
use Dyson equations for the lesser and bigger components
(\ref{dyeq}), as well as the Fourier representation of the retarded Green's function. We get:
\begin{eqnarray}
& & \hat{G}^R(t,t')  =  - i \Theta (t-t') \sum_{\alpha=1 }^{N_r}
\sum_{k_1 k_2 k_3 -\infty}^{\infty}
  \int_{-\infty}^{\infty}
\frac{d\omega}{2\pi} \hat{\Gamma}(k_2,\omega) \times
\nonumber \\
& &  e^{-i [\omega (t-t') + \Omega_0 (k_1 t - k_3 t')]}\hat{\cal G}(k_1, \omega + k_2 \Omega_0 )
  \hat{\cal G}(k_3, \omega)^{\dagger}.
\end{eqnarray}
Transforming the above Green's function according to (\ref{fourfre})
and (\ref{disfre}) allows us to write the following spectral representation:
\begin{eqnarray}
& & \hat{\cal G}(k, \omega)
= \sum_{k' k''=-\infty}^{\infty}\int_{-\infty}^{\infty} \frac{d \omega'}{2 \pi}
\nonumber \\
& & 
\times \frac{ \hat{\cal G}(k+k', \omega' + k'' \Omega_0 )
 \hat{\Gamma} (k'',\omega') \hat{\cal G}(k', \omega')^{\dagger}}
{\omega - (\omega' + k' \Omega_0 ) + i 0^+}.
\end{eqnarray}

For the case of stationary reservoirs, it is easy to prove from the above
expression the following identity:
\begin{eqnarray}
& & \hat{\cal G}(k,\omega ) -\hat{\cal G}(-k, \omega + k \Omega_0 )^{\dagger}
= -i \sum_{k'=-\infty}^{\infty} \hat{\cal G}(k+k', \omega - k' \Omega_0)
\nonumber \\
& &
\times
 \hat{\Gamma}^0 (\omega- k' \Omega_0) \hat{\cal G}(k', \omega- k' \Omega_0)^{\dagger},
\label{shoap}
\end{eqnarray}
which for $k=0$ reduces to:
\begin{eqnarray}
& & \hat{\cal G}(0,\omega ) -\hat{\cal G}(0, \omega  )^{\dagger}= 
-i \sum_{k'=-\infty}^{\infty} \hat{\cal G}(k', \omega - k' \Omega_0)
 \nonumber\\
& &
\times
 \hat{\Gamma}^0 (\omega- k' \Omega_0) \hat{\cal G}(k', \omega- k' \Omega_0)^{\dagger}.
\label{shoap0}
\end{eqnarray}
The above identities are also valid in the case of reservoirs with
oscillating voltages provided that they are described by a wide-band model
with a smooth density of states 
such that 
$\Gamma^0_{\alpha}(\omega - m \Omega_0) \sim \Gamma^0_{\alpha}(\omega )$. 
This can be easily proved by using the first of the summation formulae of appendix E.

\section{ }
Products of
Bessel functions satisfy the following summation formulas:
\begin{eqnarray}
& & \sum_{n=-\infty}^{\infty} J_{n+p}(X) J_n(X) = \delta_{p,0}, \nonumber \\
& &  \sum_{n=-\infty}^{\infty} J_{n+p}(X) J_n(X) n = X (\delta_{p,1} + \delta_{p,-1})
\nonumber \\
& & \sum_{n=-\infty}^{\infty} J_{n+p}(X) J_n(X) n^2 = \frac{X^2}{2} \delta_{p,0} 
- X \Big[ (p-\frac{1}{2}) \delta_{p,1} \nonumber \\
& &  - (p + \frac{1}{2})\delta_{p,-1} \Big]
+ \frac{X^2}{4} (\delta_{p,2} + \delta_{p,-2}).
\label{besum}
\end{eqnarray}

\end{document}